\documentclass[10pt,letterpaper]{article}
\usepackage[top=0.85in,left=1in,right=1in,footskip=0.75in]{geometry}

\usepackage{changepage}

\usepackage[utf8]{inputenc}

\usepackage{textcomp,marvosym}

\usepackage{fixltx2e}

\usepackage{amsmath,amssymb}

\usepackage{cite}

\usepackage{nameref,hyperref}

\usepackage{microtype}
\DisableLigatures[f]{encoding = *, family = * }

\usepackage{rotating}

\usepackage[aboveskip=1pt,labelfont=bf,labelsep=period,justification=raggedright,singlelinecheck=off]{caption}
\makeatletter
\renewcommand{\@biblabel}[1]{\quad#1.}
\makeatother

\usepackage{hyperref}
\usepackage{amssymb}
\usepackage{amsfonts}
\usepackage{amsmath}
\usepackage{endnotes}
\usepackage{graphicx}
\usepackage{rotating}
\usepackage{dcolumn}
\usepackage{tablefootnote}
\usepackage{xcolor}
\usepackage{cite}
\usepackage{rotating}
\usepackage{booktabs,threeparttable}
\usepackage{caption}
\usepackage[titletoc, title]{appendix}
\usepackage{amsmath}
\usepackage{graphicx}
\usepackage{subfig}
\usepackage{float}
\usepackage{mathrsfs}

\date{}

\usepackage{lastpage,fancyhdr,graphicx}
\usepackage{epstopdf}
\fancyheadoffset[L]{2.25in}
\fancyfootoffset[L]{2.25in}

\newcolumntype{d}[1]{D..{#1}}
\hypersetup{
    colorlinks,
    linkcolor={red!50!black},
    citecolor={blue!50!black},
    urlcolor={blue!80!black}
}

\makeatletter
\newcommand*{\rom}[1]{\expandafter\@slowromancap\romannumeral #1@}
\makeatother

\begin{document}

\begin{flushleft}
{\Large
\textbf\newline{The Impact of Services on Economic Complexity: Service Sophistication as Route for Economic Growth}
}
\newline
\\
Viktor Stojkoski\textsuperscript{1},
Zoran Utkovski\textsuperscript{1,3},
Ljup\v{c}o Kocarev\textsuperscript{1,2,*}
\\
\bigskip
\bf{1} Macedonian Academy of Sciences and Arts, Skopje, Macedonia
\\
\bf{2} Faculty of Computer Science and Engineering, Ss. Cyril and Methodius University Skopje, Macedonia
\\
\bf{3} Faculty of Computer Science, University Goce Del\v{c}ev \v{S}tip, Macedonia 
\bigskip


* E-mail: lkocarev@manu.edu.mk

\end{flushleft}
\section*{Abstract}
Economic complexity reflects the amount of knowledge that is embedded in the productive structure of an economy. By combining tools from network science and econometrics, a robust and stable relationship between a country's productive structure and its economic growth has been established. Here we report that not only goods but also services are important for predicting the rate at which countries will grow. By adopting a terminology which classifies manufactured goods and delivered services as products, we investigate the influence of services on the country's productive structure. In particular, we provide evidence that complexity indices for services are in general higher than those for goods, which is reflected in a general tendency to rank countries with developed service sector higher than countries with economy centred on manufacturing of goods. By focusing on country dynamics based on experimental data, we investigate the impact of services on the economic complexity of countries measured in the product space (consisting of both goods and services). Importantly, we show that diversification of service exports and its sophistication can provide an additional route for economic growth in both developing and developed countries.     

\section*{Introduction}
\label{Introduction}

The concept of \textit{economic complexity} has been recently introduced \cite{Hidalgo-2007,Hidalgo-2009,Hidalgo-2011,Hidalgo-2012,Hidalgo-2014} with the aim to reflect the amount of knowledge that is embedded in the \textit{productive structure} of an economy. By leveraging tools from network science and econometrics, it has been documented that a robust and stable relationship between a country's productive structure and its economic growth exists \cite{Hidalgo-2007,Hidalgo-2009,Hidalgo-2011,Hidalgo-2012,Hidalgo-2014,Frenken-2007,Saviotti-2008,Cristelli-2015}. Central to economic complexity is the description of the relations between countries and products they export via a bipartite network (graph), along with the thereby associated \textit{country-product} (adjacency) matrix. By using the information contained in the country-product
matrix constructed according to Balassa’s \cite{Balassa-1964}
\textit{revealed comparative advantage} (RCA), Hidalgo and Hausmann proposed a method\cite{Hidalgo-2009}, also referred to as the \textit{method of reflections} (MR), to derive \textit{complexity} of countries and products as the components of the fixed point solution of an iterative linear map. As a result, countries in the international market are ranked and the difference in their competitiveness is measured based on their complexity score. The method, similar in the spirit to pagerank algorithms, \cite{Page-1998}, is based on the hypothesis of a linear relation (more precisely an arithmetic average) between the complexity of a product, and the complexity (i.e. competitiveness) of its exporters.
 
Being motivated by the triangular structure of real data in
the country-product matrix, in \cite{Tacchella-2012, Cristelli-2013} the authors propose an 
approach which, as they argue, is more suited to reflect the ideas underlying
the arguments of a capability driven economic competitiveness. The novelty of the approach is in the introduction of \textit{fitness} as a measure for the economic complexity (i.e. competitiveness) of a country and, crucially, in the proposal of a \textit{highly nonlinear} relationship between the complexity of products
and the fitness of countries producing them. The nonlinear algorithm, also referred to as \textit{fitness-complexity method} (FCM), and the resulting nonlinear metric, are constructed on the premise that, although the competitiveness of a country is mostly determined by the diversification of its exports, the complexity of a product is mostly determined by the complexity of the least competitive exporting countries. The FCM metric has been shown to be economically-grounded \cite{Tacchella-2012, Cristelli-2013} and to be effective in ranking countries and products by their importance in the network \cite{Mariani-2015}.
While the results obtained with both methods generally support the relationship between a country's productive structure and its economic growth, they essentially introduce non-monetary and non-income-based measures for a country's economic complexity which uncover its hidden potential for development and growth.

The concept of economic complexity, as originally introduced, focuses on the role of goods exported by a country for the country's competitiveness and economic growth. While goods manufacturing and export has traditionally been assumed to be the single most important indicator for the economic development of countries, the role of services as drivers of growth has not been investigated to the same extent. As per capita income increases, most countries witness a rising share of services in the total output, accounting to over 60 percent of global GDP in the last decade. The negligence of services in the policy and research debate may stem from earlier observations that services are associated with low productivity and are merely inputs in the production of goods. However, because of the revolution in technology (in particular information and communication technology-ICT) and the rapid growth in transport and trade – with the advent of globalisation and the internet age, as argued in \cite{Mishra-2011}, "services are no longer exclusively an input for trade in goods but have become a final export, i.e. a product for direct consumption".

In view of these changes in the nature of services and their growing importance, we investigate their role in the context of economic complexity. We adopt the nomenclature according to which products encompass both goods and services, which is in line with current trends in economy where services are seen as products for direct consumption. Following this convention, here we refer to the amount of knowledge that is required to produce a good or to deliver a service as \textit{productive knowledge}. This knowledge can vary enormously from one good to the next or from one service to the next. In a previous attempt to extend the concept of economic complexity to services, Hidalgo and Hausmann \cite{Hidalgo-2014} used aggregated service data together with (highly) disaggregated goods data to evaluate the complexity of countries and products. For the reason that service data was still not disaggregate enough to be included in the complexity calculations, they quickly dismissed the service inclusion. In order to circumvent the main problem associated with the inclusion of services in the model (which is the lack of highly disaggregated data on services, as opposed to goods), here we take a different path and aggregate goods data (to make it comparable to the service data), resulting in \textit{aggregated} indices (i.e. indices based on aggregated data) as measures for the complexity of products and countries. We thereby show that, although some information is inevitably lost in the process of aggregation, the inclusion of services in the model provides new insights for countries and products. 

With the aim to investigate the impact of services on the countries' productive structure, we present results obtained by using both the linear method (MR) due to Hidalgo and Hausmann, and the nonlinear method (FCM)  due to Tachella et al. In addition, we address the important issue of convergence of nonlinear algorithms when applied to \textit{nested networks} (such as the bipartite contry-product network). The results of our investigation, motivated by the recent reports \cite{Pugliese-2014} and \cite{Wu-2016} which relate the convergence of the FCM fitness and complexity scores to the shape of the underlying nested country-product matrix, suggest that country-product matrices obtained from real trade data often exhibit an "unfavorable" structure resulting in some country fitness and product complexity scores converging to zero. As a consequence, the stability of the induced country and product rankings may be also affected. Motivated by these findings, we propose a new nonlinear method (which may be seen as a modification of the fitness-complexity method), with tractable convergence. In the following we refer to it as the modified-fitness complexity method (M-FCM). We thereby show that with the introduced modification, fitness and complexity scores converge to values strictly greater than zero, resulting in more stable country and product rankings. The details about the proposed metric and the convergence of the algorithm are presented in the \nameref{Materials and Methods} section.
      
By performing an analysis of the new aggregated indices (obtained from aggregated goods and service data) with both MR and M-FCM, we report that not only goods but also services are important for explaining economic stability and growth. In fact, we provide evidence that complexity indices for services are in general higher than those for goods, thus putting most services above in the \textit{product} complexity ranking. As a result, countries with developed service sector rank in general higher on the scale of aggregated economic complexity indices, as compared to the case when ranking is performed based on disaggregated goods data only. While there remains the non-monetary and non-income-based dimension of the productive knowledge as hidden potential for development and growth, we also show that there is a relation between the economic complexity of a country measured in the product space (including goods and services) and the income per capita that the country is able to generate measured with GDP per-capita adjusted for Purchasing Power Parity (PPP). Importantly, our results suggest that growth in service exports and its sophistication can provide an additional route for economic growth in both developing and developed countries. We argue that the converse may also hold - the countries (including those with diversified goods portfolio, i.e. saturated goods space) which are not able to accordingly populate the service space (i.e. diversify the service portfolio), may face diminishing growth prospects. These conclusions, which can not be inferred from the analysis based solely on goods data, may be more consistent with the "perceived" economic situation in some countries than otherwise suggested by the results in \cite{Hidalgo-2009,Hidalgo-2014} and \cite{Tacchella-2012,Cristelli-2013}. 

This paper is organized as follows. \nameref{Materials and Methods} section starts by describing the trade data and their adjustment for adequate comparison of goods and services. We proceed by summarizing the method of reflections (MR) (and the induced linear metric) due to Hidalgo and Hausmann \cite{Hidalgo-2009}, together with the nonlinear fitness-complexity method (FCM) (and the induced fitness-complexity metric) developed by Tacchella et al.\cite{Tacchella-2012}. We also discuss the issues related to the convergence of the fitness-complexity algorithm and the consequences of its' application to aggregated product-country data. Here, we introduce the proposed modification of the fitness complexity method (M-FCM) and study the convergence of the resulting country fitness and product complexity scores. The \nameref{Results} section consists of five parts. In the first part we discuss the implications of data aggregation and inclusion of services to the model. In the second part, we proceed by estimating few baseline regressions of growth on our aggregated complexity measures which include both goods and services, and by comparing them with standard (disaggregated) complexity measures, which account for goods only, and the simplest complexity measures that can be derived from the country-product network. This allows us to study whether the aggregation of goods data and the inclusion of services still yields country complexity/fitness indices that correlate significantly with future growth. In the third part we look at something that the branch of Economic Complexity has widely eschewed until now - the differences between the complexity of services and goods. We do this by comparing services and goods rankings obtained through both linear and nonlinear methods. The fourth part of this section examines the similarities between the aggregated and disaggregated complexity measures. Finally, in the last (fifth) part of this section we present insights on the additional information that our aggregated metrics carry for the economic potential of the countries, by providing examples for selected countries. The \hyperref[Conclusion]{last} section concludes our findings and gives suggestions for future research.

\section*{Materials and Methods}
\label{Materials and Methods}

\subsection*{Data}
\label{Data}

To analyse the association between countries and products we collect international trade data from various sources. On the one hand, there are not many service datasets; even those that exist do not provide quality data. The best one that can be found is the World Bank Trade in Services Database \cite{Francois-2013} based on the Extended Balance of Payments Classification. The creators of this dataset consolidate multiple sources of bilateral trade data in services and calculate the export trade flows of a reporter by using information on imports of the partner country. By doing so, they identify and correct most of the inconsistencies present in the individual sources for service trade data. Nevertheless, since the nature of the trade in services makes it difficult to collect service data as accurately as goods data, it includes data mainly on two out of the four modes of trade in services; cross-border trade flows and consumption abroad. A detailed description of the construction of the dataset is provided in Appendix B of \cite{Saez-2014}.
	
The lowest aggregation level that can be used appropriately is the 3-digit level. With this classification, we recognize 12 different types of services: 11 being the standard service components in the IMF Balance of Payments Manual, whereas the last service type that we include is EBOPS 983: Services not allocated category, which serves as an approximation for all other services that could not be allocated in any of the standard categories. We apply only one type of data sanitation, in the case when there are disaggregated country trade flows, but no aggregated ones. Then, we sum the disaggregated trade flow and use it as a proxy of the 3-digit level service trade flow of the country.

Compared to service data, goods data is abundant and finely dissagregated. For the period of 1988-2000 we use the Feenstra et al.\cite{Feenstra-2005} dataset from the Center for International Data which disaggregates the goods to SITC rev.2  to 4-digit level. The dataset itself is an advanced version of the UN COMTRADE database as it cleans the data from it in the same way as the Trade in Services Database does. We extend the dataset until 2010 by using the trade data provided by UN COMTRADE\cite{UNCOMTRADE}. To make goods data comparable to service data, we aggregate it to the Standard International Trade Classification(SITC) rev.2 1-digit level, thus ending with 10 classes of goods. The resulting goods and services are presented in Table \ref{Table:PandS}. For further cleaning of unreliable or inadequately classified data, we restrict the dataset to the same countries that were used in the Atlas of Economic Complexity \cite{Hidalgo-2014,Simoes-2010}.

\begin{table} [b!]
\begin{adjustwidth}{0in}{0in}
\caption{\textbf{The included products and services}} 
\centering
\setlength{\tabcolsep}{0.1cm}
\begin{tabular}{|c|l|c|l|} 
\hline
\textbf{SITC} &\textbf{Good} &\textbf{EBOPS} &\textbf{Service} \\
\hline
0 & Food and live Animals & 205 & Transportation \\\hline
1 & Beverages and Tobacco & 236 & Travel \\\hline
2 & Crude materials, inedible, except fuels & 245 & Communication \\\hline
3 & Mineral fuels, lubricants and related materials & 249 &Construcion\\\hline
4 & Animal and vegetable oils, fats and waxes & 253 & Insurance \\\hline
5 & Chemicals and related products, n.e.s. & 260 & Financial Services \\\hline
6 & Manufactured goods  & 262 & Computer and information \\\hline
7 & Machinery and transport equipment & 266 & Royalties and License fees \\\hline
8 & Miscellaneous manufactured articles & 268 & Other business services \\\hline
9 & Commodities not classified elsewhere & 287 & Personal
services \\\hline
&& 291 & Government services \\\hline
&& 983 & Services not allocated \\\hline
\end{tabular}
\label{Table:PandS}
\end{adjustwidth}
\end{table}
Even though that through this aggregated classification we end up with only 22 broad categories of products (10 categories of goods and 12 categories of services) and a total of 130 countries, we believe that they are detailed enough to provide new interesting facts about the complexity of goods, services and the countries exporting them. Therefore, we utilize this classification to calculate our aggregated complexity measures.

For the baseline regression analysis of growth on the aggregated complexity metrics we use data on GDP per capita at PPP in constant 2005 dollars, service value added, exports and population from the World Bank World Development Indicators Database, whereas data from the Observatory of Economic Complexity \cite{Simoes-2010} is taken to individually calculate the Complexity Outlook Gain. These data are publicly available at \url{https://dx.doi.org/10.6084/m9.figshare.3404728.v1}.

\subsection*{Methodology}

\paragraph*{Bipartite Network of Countries and Products:}
Following Hidalgo and Hausmann\cite{Hidalgo-2007,Hidalgo-2009,Hidalgo-2011,Hidalgo-2012,Hidalgo-2014}, based on the collected data, we construct a bipartite network in which countries are connected to the products (here both goods and services) they export. According to this representation, the existance of a link between a country and a product in the bipartite network is related to the revealed comparative advantage (RCA) index introduced by Balassa \cite{Balassa-1964}. RCA is widely used in international economics for calculating the relative advantage or disadvantage of a certain country in a certain class of goods or services as evidenced by trade flows. Formally, given a set of $\mathscr{C}$ countries and $\mathscr{P}$ products, a set of RCA indices relating the countries and the products they export, is defined as: 
\begin{equation} \label{eq:RCA}
RCA_{ij} = \frac{E_{ij}/\sum_j E_{ij}}{\sum_i E_{ij}/\sum_{i,j} E_{ij}} 
\end{equation}
where $E_{ij}$ is the export of product $j$ by country $i$, with $i\in [\mathscr{C}]$ and  $j\in [\mathscr{P}]$ (where $[\mathscr{K}]$ denotes the set $\{1,2,\ldots,\mathscr{K}\}$). As such, $RCA_{ij}$ indicates whether the country $i$ is significant exporter product $j$ or not. For example, $RCA_{ij}>1$ indicates that country $i$'s share of product $j$ (good or service) is larger than the product's share of the entire world market, thus "revealing" a comparative advantage of country $i$ with respect to product $j$. 

The entries of the country-product (adjacency) matrix $M$ describing the links between countries and products in the bipartite network (country-product space) may be related to the corresponding RCA index in different ways. In the original representation according to \cite{Hidalgo-2009,Hidalgo-2014}, the $M_{ij}$ entry of the adjacency matrix is described as:
\begin{equation} \label{eq:mentry}
 M_{ij}= 
\begin{cases}
1,& \text{if}\ RCA_{ij} \geq 1 \\\
0,              & \text{otherwise}
\end{cases}
\end{equation}
When working with aggregated goods and service data, however, one may argue that computing a unique RCA matrix by joining services and goods in one set of products may produce several biases, due to their different nature and magnitude (volume) of trade. Although the concept of RCA already incorporates these differences to a certain extent (by the normalization taking place in (\ref{eq:RCA})), in the \nameref{S1} we also discuss an alternative RCA representation, where the RCA indices for goods and services are estimated separately (by creating two different sets, one of goods and another of services), and are then concatenated into a combined RCA matrix. Nonetheless, as we will see later on, the statistical results that we provide in S1 suggest that the aggregated measures obtained through the representation with a unique RCA matrix tend to be more appropriate, as they provide an economically more consistent interpretation of the countries' productive structure.

\paragraph*{Complexity of Countries and Products:}
The observations about the structure of the country-product matrix have motivated a series of works, including \cite{Hidalgo-2009,Hidalgo-2014} and \cite{Tacchella-2012, Cristelli-2013} which interpret economic complexity by special fundamental endowments, called \textit{capabilities}. According to this interpretation, capabilities represent "the mixture of all available resources in a given economy and the features of the national social organization which make possible the production and export of tradable products." As the authors in \cite{Tacchella-2012, Cristelli-2013} argue, capabilities may also be seen as "intangible assets which drive the development, the wealth and the competitiveness of a country". In practice they determine the complexity of a productive system as each product requires a specific set of necessary capabilities which must be owned by a country in order to produce it and then to export it. 

Due to the difficulty in categorizing and analysing capabilities quantitatively, exported products by each country become in such a scenario the main proxy to infer the endowment of capabilities, i.e. the level of complexity of a productive system. In this context, two simple measures may be introduced. The first measure, called \textit{diversity}, is the country's degree in the bipartite network, which for country $i$ reads:
\begin{align}  \label{eq:div}
d_{i}= \sum_{j} M_{ij}.
\end{align} 
The second measure, associated with products, is called \textit{ubiquity}. For a product $j$, ubiquity is defined by the number of countries exporting it:
\begin{align} \label{eq:ubq}
u_{j}= \sum_{i} M_{ij}
\end{align}
Diversity and ubiquity are inversely related: higher diversity means that a country has an export basket with many different products and, hence, it has a high amount of know-how; higher ubiquity means that a product (good or a service) is included in many countries' export baskets, and thus it needs fewer capabilities to be produced. However, both diversity and ubiquity are simple graph characteristics (node degree) of the bipartite network represented by the adjacency matrix $M$ which carry limited information about the productive structure of a country or complexity of a product as they do not take into account \emph{who} else exports the same products. As a result, a careful assessment is required if any of these simple measures is to be used for the explanation of economic phenomena. 

For these reasons, two approaches which extend over the concepts of diversity and ubiquity have been introduced in the literature, the details of which we present in the following.

\paragraph*{The Method of Reflections (MR):}
The first approach, due to Hidalgo and Hausmann \cite{Hidalgo-2009,Hidalgo-2014}, is organised around a linear method, implicitly constructed on the premise that the number of a country’s capabilities is equal to the average number of capabilities required by its exporting products, whereas the number of capabilities required by a product is the average number of capabilities present in the countries that are exporting it. The method is describes an iterative procedure with the aim to obtain measures for the complexity of countries and products. According to the algorithm, the complexity scores of the country $i$, respectively product $j$, as calculated after $n$ iterations, are given by:
\begin{equation} 
\begin{cases}
 c_{i,n}= \frac{1}{d_{i}} \sum_{j} M_{ij} p_{j,n-1}  \\\
 
 p_{j,n}= \frac{1}{u_{j}} \sum_{i} M_{ij} c_{i,n-1}
\end{cases}
\label{eq:MR}
\end{equation}
where the initial conditions are $c_{i,0}=d_{i}$ and $p_{j,0}=u_{j}$, with $d_{i}$ and $u_{j}$ denoting diversity, respectively ubiquity, as defined in (\ref{eq:div}), respectively (\ref{eq:ubq}). By inserting the product (country) complexity relation into the country (product) complexity, one can define $c_{i,n}$ ($p_{j,n}$) as a linear transformation (map) of $c_{i,n-2}$ ($p_{j,n-2}$). When written in vector form, the linear operator characterizing this linear map represents the transpose of an ergodic Markov transition operator \cite{Cristelli-2013}, implying that all $c_{i,n}$ and $p_{j,n}$ converge to a same constant with speed of convergence proportional to the second largest right eigenvalue. For more details about the method of refelections and its' properties, please see \cite{Hidalgo-2009, Cristelli-2013, Caldarelli-2012} and references therein.

It is important to note that the iterations can be evaded by generating two new matrices for countries and products. Then, the capability measures can be obtained through ranking the nodes in each of these networks by calculating the second largest eigenvectors of the matrices. Formally, for countries, define the matrix $[\tilde{C}_{ii'}]$ as
\begin{align*}
\tilde{C}_{ii'} = \sum_j \frac{M_{ij} M_{i'j}}{d_{i} u_{j}}
\end{align*}
The resulting country complexity measure, called \textit{economic complexity index (ECI)}, is related to the normalized eigenvector associated with the second largest eigenvalue of $\tilde{C}$ 
\begin{align*}
ECI= \frac{\vec{c} -  \langle \vec{c} \rangle }{ \mbox{stdev} (\vec{c})},
\end{align*}
where $ \langle \cdot \rangle  $ represents an average, $\mbox{stdev}(\cdot)$ stands for the standard deviation, and $\vec{c} $ is the eigenvector associated with the second largest eigenvalue of $\tilde{C} = [\tilde{C}_{ii'}]$. 

By analogy, the resulting linear product complexity measure is named \textit{product complexity index} (PCI) and is defined by substituting the index of countries $(i)$ with that for products $(j)$ in the definitions above. In other words, the linear PCI is obtained by analyzing the matrix connecting product $j$ to product $j'$, according to the number of countries that export both products:
\begin{align*}
\tilde{P}_{jj'} = \sum_i \frac{M_{ij} M_{ij'}}{d_{i} u_{j}}
\end{align*}	
Formally, PCI is given by
\begin{align*}
PCI = \frac{\vec{p} -  \langle \vec{p} \rangle }{ \mbox{stdev} (\vec{p})}
\end{align*}
where $\vec{p}$ is the eigenvector of $\tilde{M}_{jj'}$ associated with the second largest eigenvalue.

We implement this method to our aggregated representation of products to estimate the corresponding \textit{aggregated} ECI and \textit{aggregated} linear PCI.

\paragraph*{The Fitness-Complexity Method (FCM):}
The second approach we address, introduced in \cite{Tacchella-2012, Cristelli-2013}, is similar in respect of its' aim to the approach due to Hidalgo and Hausmann. At the focus there is the concept of \textit{fitness} which, similar to ECI, serves as a measure for a country's competitiveness. This method is based on an iterative procedure which couples the fitness of a country to the complexity of the products it exports. While fitness, similarly to country complexity (i.e. ECI index) in MR, is defined to be proportional to the (weighted) sum of the complexities of the exported products, the complexity of a product, in contrast to MR, is now no longer defined as the average fitness of the countries producing it. Instead, a strong nonlinear relationship between the complexity of an exported product and the competitiveness of its producers is assumed. The nonlinear map is formally described by the following set of (iterative) equations:
\begin{equation}
\begin{cases}
\widetilde{c}_{i,n} = \sum_j M_{ij} p_{j,n-1} \\\
    
\widetilde{p}_{j,n}= \frac{1}{\sum_i M_{ij} \frac{1}{c_{i,n-1}}}
\end{cases}
\longrightarrow	
\begin{cases}    
c_{i,n} = \frac{\widetilde{c}_{i,n}}{\langle\widetilde{c}_{i,n}\rangle} \\\

p_{j,n} = \frac{\widetilde{p}_{j,n}}{\langle\widetilde{p}_{j,n}\rangle},
\end{cases}
\label{eq:FCM}
\end{equation}
where $\widetilde{c}_{i,n}$ and $\widetilde{p}_{j,n}$ are, respectively, the intermediate values of the fitness (i.e. complexity) of the country $i$ and the (nonlinear) complexity index of a product $j$, as calculated after $n$ iterations of the algorithm. After each step, the intermediate values are normalized to $c_{i,n}$ and $p_{j,n}$ by separately dividing them with the corresponding averages over all countries and products, i.e. $\langle\widetilde{c}_{i,n}\rangle$ and ${\langle\widetilde{p}_{j,n}\rangle}$. Different to MR, here the initial conditions are given by $\widetilde{c}_{i,0}=1$ and $\widetilde{p}_{j,0}=1$, for all $i$ and $j$. 

Depending on the possible values that the elements of the $M$ matrix take ( binary vs. real-valued), two different definitions of fitness are defined in \cite{Tacchella-2012,Cristelli-2013}: the first one, denoted as  \textit{intensive fitness} (IF), is obtained from the binary representation of $M$, as presented in (\ref{eq:mentry}); the second one, denoted as \textit{extensive fitness} (EF), is based on a non-binary (i.e. weighted representation) of the elements of the country-product matrix $M$. While there are instances where the use of EF over IF may be beneficial, EF is, by construction, more appropriate for capturing the short-term competitiveness of a nation, rather than its' long-term competitiveness and robustness of the productive structure \cite{Tacchella-2012, Cristelli-2013} . For these reasons, in our analysis we will concentrate on the intensive fitness, with the remark that the results easily generalize to the case when the extensive fitness is used as a complexity measure. 

An attractive property of the fitness-complexity method is that the resulting  metrics converges to a non-trivial unique fixed point after several iterations. However, it may happen that in the steady state the complexity (fitness) indices $c_{i}$ for some of the countries converge to zero which, due to the nonlinear relation, implies convergence to zero of the complexity indices $p_{j}$ of some of their exported products \cite{Pugliese-2014, Wu-2016}. In general, this occurrence may be seen as a drawback of the nonlinear fitness-complexity method since, in that case, it fails to compute any differences among certain countries and products. Whether there will be countries and products whose complexity scores converges to zero, depends solely on the shape of the country-product matrix $M$, i.e. on the structure of the trade data (see \cite{Pugliese-2014} for more detailed discussion). In line with this observation, the results of our investigation with real data show that in most of the years the matrices that we construct from aggregated goods and service data have, indeed, an "unfavorable" shape (according to the findings in \cite{Pugliese-2014, Wu-2016}), resulting in some of the complexity indices for countries and products converging to zero. Hence, the direct application of FCM to our dataset may not be appropriate, requiring an alternative approach. In order to deal with this weakness, and, consequently, to accommodate FCM to our type of data, in the following we propose a modification of the method.

\paragraph*{Modified-Fitness-Complexity Method: (M-FCM)} 
Here we define a modification of the nonlinear Fitness-Complexity Method as follows:
\begin{equation}
\begin{cases}
    \widetilde{c}_{i,n} = \sum_j M_{ij} p_{j,n-1} \\\
    
    \widetilde{p}_{j,n}= \frac{1}{\sum_i M_{ij} \left( N_c-{c_{i,n-1}} \right)}
	\end{cases}
\longrightarrow	
 \begin{cases}    
c_{i,n} = \frac{\widetilde{c}_{i,n}}{\langle\widetilde{c}_{i,n}\rangle} \\\

p_{j,n} = \frac{\widetilde{p}_{j,n}}{\langle\widetilde{p}_{j,n}\rangle}
\end{cases}
\label{eq:M-FCM}
\end{equation}
The difference with the original nonlinear method is in the equation for $\widetilde{p}_{j,n}$ where we substitute the term $\frac{1}{c_{i,n-1}}$ with $ \left( N_c-{c_{i,n-1}} \right)$,  where $N_c$ is the number of countries (the number of rows in the $M$ matrix). We argue that, with this modification, the main flavour of the fitness-complexity method is still preserved. Indeed, here, as in the original method, the complexity of a product is still (mostly) determined by the complexity of the least competitive exporting countries.

As a matter of fact, our proposed modification may be seen as a second order approximation of the original nonlinear method since the Taylor series expansion of $1/c_{i,n-1}$ at the point $N_c/2$ is proportional to $(N_c-c_{i,n-1})$: 
\begin{align*}
\frac{1}{c_{i,n-1}}&=\frac{2}{N_c}-\frac{4(c_{i,n-1}-\frac{N_c}{2})}{N_c^2}+O((c_{i,n-1}-\frac{N_c}{2})^2)\nonumber\\ &\approx \frac{4}{N_c^2} (N_c-c_{i,n-1}) \propto (N_c-c_{i,n-1}),
\end{align*}
which converges as long as $|N_c-c_{i,n-1}|<N_c$.

Yet, the introduced modification has an additional fine property as it yields  complexity scores (for both countries and products) that always converge to strictly positive, finite values. In order to prove this property, we write the steady-state solutions of $c_i$, $\widetilde{c}_i$, $p_j$, and $\widetilde{p}_j$:
\begin{align}
\widetilde{c}_{i} &=\sum_j M_{ij} p_{j};  \label{eq1} \\ 
\widetilde{p}_{j} &=  \frac{1}{\sum_i M_{ij} \left( N_c - c_{i} \right); } \label{eq2} \\
c_{i} & = \frac{\widetilde{c}_{i}}{\langle\widetilde{c}_{i}\rangle_i}; \label{eq3} \\
p_{j} &= \frac{\widetilde{p}_{j}}{\langle\widetilde{p}_{j}\rangle_j}. \label{eq4}
\end{align}
It is easy to see, starting from the initial conditions that $\widetilde{c}_{i,n} > 0$, $\widetilde{p}_{j,n}>0$,  ${c}_{i,n} > 0$, and ${p}_{j,n}>0$. Therefore, 
 $\widetilde{c}_{i} \geq 0$, $\widetilde{p}_j \geq 0$,  ${c}_{i} \geq 0$, and $p_j \geq 0$.
Note that due to the normalization process, after each step we have $N_c= \sum_i c_{i,n} = \sum_i c_i$ for each $n$, and the total number of products in the network (the number of columns in the $M$ matrix) is $N_p = \sum_j p_{j,n} = \sum_j p_j $ for each $n$. We assume that each country exports at least one product ($d_i =  \sum_{j} M_{ij} >0$) and each product is exported by at least one country ($u_j = \sum_{i} M_{ij}  >0$). 

From $\sum_i M_{ij}c_i \geq 0$ it follows that 
\begin{equation*}
\sum_i M_{ij}(N_c-{c_{i}})  = N_c u_j - \sum_i M_{ij} c_{i} \leq N_c u_j 
\end{equation*}
Hence 
\begin{equation} \label{eq-gre-0}
\frac{1}{ \sum_i M_{ij}(N_c-{c_{i}})} \geq \frac{1}{N_c u_j} > 0
\end{equation}
since $u_j>0$. Therefore, from Eqs~(\ref{eq1}), (\ref{eq2}), (\ref{eq3}), and (\ref{eq4}) we have 
 $\widetilde{c}_{i} > 0$, $\widetilde{p}_j > 0$,  ${c}_{i} > 0$, and $p_j> 0$.
From  (\ref{eq-gre-0}), and since $\sum_i M_{ij} >0$ and $N_c=\sum_i c_i$, we have $c_i < N_c$. Thus steady-state solutions are also finite. 

Due to these favorable characteristics of M-FCM, we will apply it instead of the originally proposed FCM for the computation of the fitness-complexity measures based on aggregated data. We will refer to the resulting indices as  \textit{aggregated IF} and \textit{aggregated nonlinear PCI}. Additionally, we will also use M-FCM to compute the complexity measures for disaggregated goods data, a step which is required in the process of evaluation of the impact that services and data aggregation have on the country productive structure.

\section*{Results and Discussion}
\label{Results}
\subsection*{Data Aggregation and Inclusion of Services: Implications}
Before we proceed with a detailed analysis of the characteristics of the aggregated complexity measures and present country and product rankings based on these measures, in this part we provide an initial intuition about how the aggregation of goods and the inclusion of services may affect economic complexity indicators and country ranking. 

\paragraph*{Data aggregation:}
Intuitively, as data aggregation results in reduction of the number of products (now 22 in total, including services), any change in the coefficients of the adjacency matrix (i.e. introduction/deletion of a link in the bipartite network), affects the overall structure more significantly than in the case of the original model with 774 disaggregated goods. Indeed, the process of goods aggregation leads to distortions in the distribution of diversity among countries. The impact of this effect is evident when diversity of countries based on aggregated goods is compared to diversity obtained on the basis on disaggregated goods (services are excluded to quantify the effect of goods aggregation only). For example, in the SITC 4-digit classification developed countries such as Germany, Italy and the United States have the most diverse export basket, whereas in the SITC 1-digit classification they are replaced in the rankings with not as developed countries. While this may be interpreted as a disadvantage of goods aggregation, we note that diversity is still only a first order approximation of country complexity. Nevertheless, when services are added to the picture, the overall aggregated diversity profile resembles a lot more the diversity profile with disaggregated goods, as the diversity of developed countries is increased by the inclusion of services in the model. As in the case with disaggregated goods, countries exporting primarily natural resources are also the least diversified in this new model of aggregated products. Interestingly, when it comes to ubiquity, the above effect is less pronounced, as goods aggregation yields more predictable results with respect to the ubiquity scores of the newly obtained aggregated goods. For example, chemicals and machinery are the least ubiquitous in the aggregated goods classification (when services are excluded), which is expected having in mind that in the disaggregated classification, goods coming from these classes are among the least ubiquitous in general. When services are included the distribution of goods ubiquity remains relatively unchanged, with the only difference that each aggregated good becomes rather ubiquitous at the expense of services. Now, in the overall model of aggregated products, services such as Royalties and Finance are least ubiquitously exported in almost each year.

With a smaller adjacency matrix (i.e. a smaller number of products) as a result of aggregation, country rankings are also expected to be affected. Indeed, small changes in the RCA index may result in a transition from $0$ to $1$ in the adjacency matrix (or vice verse). As a consequence, country rankings based on both aggregated ECI and aggregated IF are more volatile (susceptible to higher variations over time) when goods are aggregated, compared to the original findings in \cite{Hidalgo-2009, Hidalgo-2014} (for ECI), and in \cite{Tacchella-2012, Cristelli-2013} (for IF) based on disaggregated goods data. This observation is consistent with the results presented later in this section, where the dynamics of country rankings is analyzed over time.   
 
Definitely, in the process of aggregation, some of the information revealed by the analyses based on disaggregated goods data is inevitably lost since some countries lose the revealed comparative advantage that they previously had in some goods (when goods were disaggregated), which is reflected in more profound changes in their overall export basket structure and their rankings. While this may be seen as a weakness of the model relying on aggregated goods data, we recall the discussion that, in the absence of finer (i.e. disaggregated service data), the most plausible way to include services in the model is to aggregate goods data to a level comparable with available service data. 

\begin{figure}[t!]
\begin{adjustwidth}{0in}{0in}
\includegraphics[width=16cm]{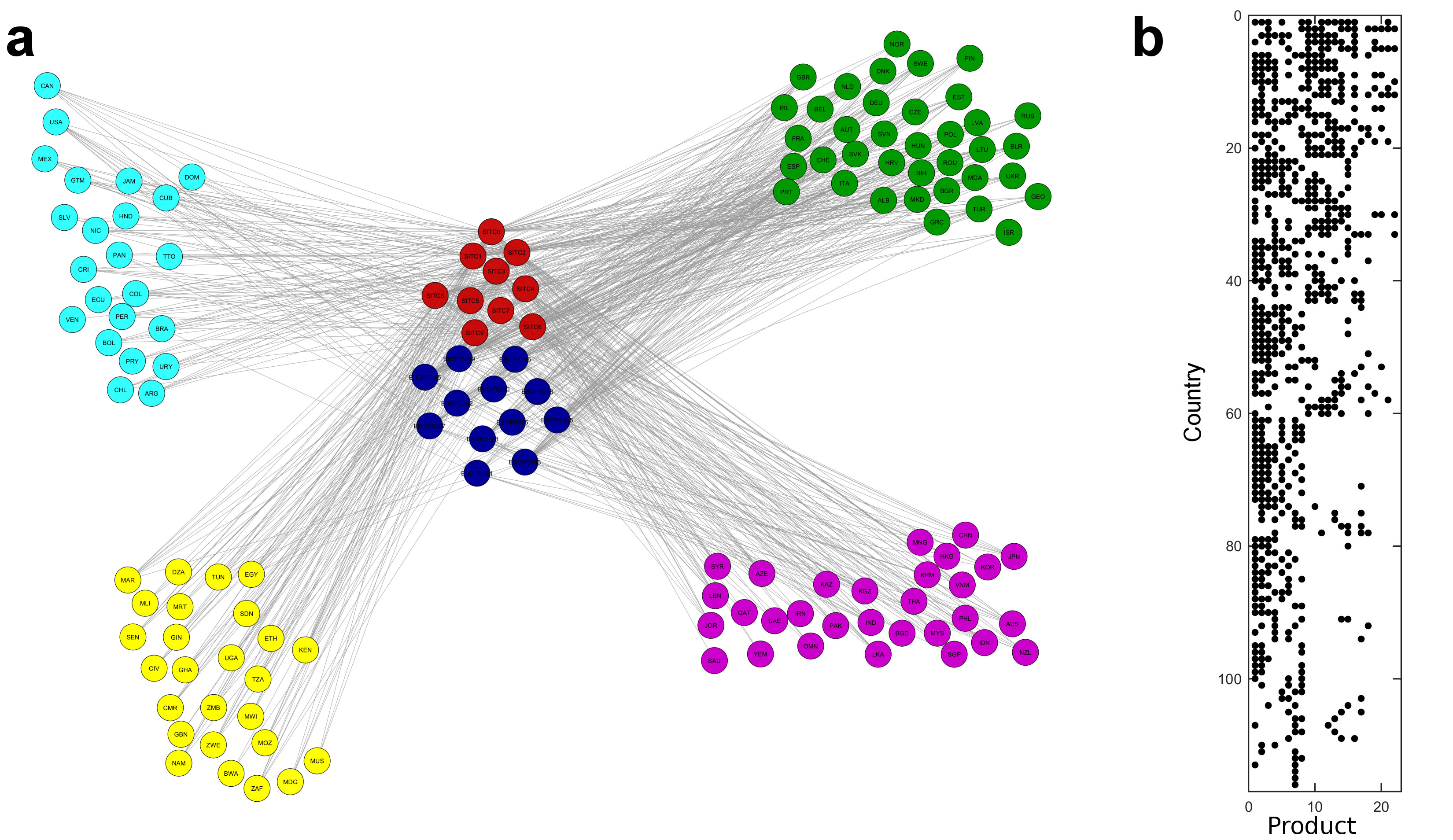}
\caption{\textbf{The Aggregated Country-Product Network}. \textbf{a)} Visualizes the country-product network for 2005. Services are given in blue and goods in red. Country nodes are colored according to the continents they belong, Europe is in green, Asia in purple, Africa in yellow, and nodes representing American countries are in cyan. \textbf{b)} Shows the $M$ matrix. Here rows represent countries, ordered accordingly to their diversity, while columns represent products  ordered according their ubiquity. \label{graph}}
\end{adjustwidth}
\end{figure}

\paragraph*{The role of services:} 
To illustrate the effect of the inclusion of services in the model, in Fig. \ref{graph}a we present a simple visualization of the country-product network for the year 2005.  

According to this representation, goods and services are located at the center and countries are grouped into regions (continents). The network consists of 116 countries, 10 products, 12 services, 448 links connecting countries and goods, and 238 links connecting countries and services when $RCA_{ij} > 1$. Initially, the difference between the number of connections of services and goods suggests that there is a significant margin between their complexity. This is also supported by the information contained in the binary $M$ matrix for the same year after rows and columns are sorted according to decreasing diversity and ubiquity, as displayed in Fig. \ref{graph}b. We observe that the matrix preserves the triangular structure present in \cite{Hidalgo-2011}, thus implying that ubiquity and diversity are also inversely related in the model with aggregated goods and services. A more thorough examination shows that, among the countries, United Kingdom and the Netherlands have the highest diversity (13), while Algeria, Saudi Arabia and Venezuela are least diversified with comparative advantage in only 1 product (Mineral fuels). The least ubiquitously exported products are Financial services, while Food and live animals chiefly for food are mostly exported by the countries. 

By separating the diversity scores of services and goods, we discover additional information about the difference in the productive capabilities embedded in services and goods. Developed and highly diversified countries, such as United Kingdom (10), Switzerland (9) and the United States (9) have the most diverse service export baskets, whereas not as developed and diversified countries, specifically Kenya (8) and Lebanon (8), have the highest goods diversity. Even though, this diversity analysis should be taken with caution since diversity alone is only a first order approximation of country complexity (fitness) and, as such, may not very informative for evaluating the complexity of a productive system, the results can serve as an indication for post-industrialization \cite{Bell-1974} - the process in which developed economies undergo a change from relying on production of goods to provision of services. Indeed, as we will see later in this section, complexity rankings performed on the basis of aggregated goods data recognize even bigger differences between service-oriented and goods-oriented developed economies.

Altogether, the effects of data aggregation and inclusion of services significantly impact the picture of the productive structure of the countries. In the following we will further investigate the impact of services on the productive structure and will discuss how their addition in the model may help us extract new valuable information about both the long-term and the short-term economic situation of nations.

\subsection*{Economic Complexity as Indicator of Future Growth}
\label{Growth Regressions}

The significance of economic complexity stems from the observation that the productive capabilities possessed by a country carry information about its long term potential. Ergo, the information contained in the adjacency matrix describing the bipartite country-product network, may be used to predict country's future growth. In this context, the performance of the method of reflections has been extensively evaluated through econometric tools in \cite{Hidalgo-2009, Hidalgo-2014}. Similarly, the proponents of the fitness-complexity method also argue that the economic complexity indicators influence economic growth \cite{Tacchella-2012, Cristelli-2013}. In order to resolve inconsistencies appearing when standard regressions are used for growth prediction, in \cite{Cristelli-2015} it is debated that there is a heterogeneous pattern of evolution of country dynamics in the fitness-income plane. Depending on whether a country has a lower or higher level of income compared to expectations based on its level of fitness, two regimes are observed which correspond to very different predictability features for the evolution of countries: in the former regime one finds a strong predictable pattern, while the second regime exhibits a very low predictability. As a result, the authors propose a selective predictability scheme which incorporates concepts from dynamical systems to deal with the heterogeneity of the evolution pattern. 

According to these observations, we argue that one may position the complexity approach in the middle between two economic concepts - one which traditionally describes economic systems in purely GDP-oriented terms, and another which substitutes GDP with new economic indicators \cite{Costanza-2009, Costanza-2014}. With this respect, we favor the stand presented in \cite{Cristelli-2015} according to which, instead of simply substituting monetary based information (such as GDP) with new, alternative indicators \cite{Costanza-2009, Costanza-2014}, it is more appropriate to complement GDP-based measures with new dimensions. The role of GDP thus remains important in our analysis, with the remark that the results should be interpreted based on the understanding that the country GDP (respectievly GDP per capita) is only a part of the picture or, more precisely, only one of the parameters in the multi-dimensional model.

While regression analysis may not be appropriate for prediction of future growth, it still provides a simple solution for inferring the validity of the expected positive relationship between growth and the aggregated complexity measures. In particular, we believe that if complexity measures are supposed to provide information about some dependent variable (i.e. long-term growth), then they should be significant and robust explanatory variables in any correct econometric model. Therefore, we stick to the regression analysis, and construct three competing econometric models in which the long term growth, controlled for possible other effects, is regressed on the aggregated and disaggregated complexity measures, as well as on the aggregated diversity (the regressions with the disaggregated ECI can be seen as a partial reproductions of the growth regressions presented in \cite{Hidalgo-2014}). We demonstrate that the aggregation of goods and the addition of services in the model still produces complexity/fitness indices which act as robust explanatory variables of growth. In addition, the regression results indicate that the obtained indices provide better approximation of the variations in the long term growth than aggregated diversity.

In the following we provide details for the analysis performed under the linear (MR) model given by (\ref{eq:MR}), and under the nonlinear (FCM) model (more precisely the modification thereof, M-FCM), as given by (\ref{eq:M-FCM}), while in \nameref{S1} we assess the robustness of our results. There, we add the initial export of goods and services (as a percent of GDP), the initial population, and the initial value added of services (again, as a percent of GDP) in the regression model. Each of these is related to both economic complexity and growth, and therefore they can be interpreted as potentially omitted variables. As such, these variables can serve for assessing the robustness of the regression analysis. Nonetheless, as shown in \nameref{S1}, in all of the addressed cases, the aggregated complexity measures remain significant explanatory variables of growth, thus removing the suspect of present bias.
 
\paragraph*{MR Regressions:}
We examine the relationship between the aggregated complexity measures and long term economic growth by estimating the following panel regression with fixed time effects:
\begin{align*}
g_{i,t}= \alpha +\beta c_{i,t}+\phi' X_{i,t}+\eta_{t}+\epsilon_{i,t}
\end{align*}
where $t=1988, 1998$ is period notation. 
The dependent variable $g_{i,t}$ is measured as the ten-year GDP (PPP) per capita growth of country $i$, between 1988-1998 and 1998-2008, i.e $g_{i,t}= \log{\left(\frac{GDP_{i,t+10}}{GDP_{i,t}}\right)}$. We focus on this period as it is longest for which service data is available. Moreover, this way we are consistent with the 'original' regression analysis performed in \cite{Hidalgo-2014}, thus allowing for direct comparison. The coefficient $\beta$ is of particular interest to us as it is an estimate of the marginal effect of $c_{i,t}$, the economic complexity of the country at the initial time $t$; $X_{i,t}$ is a vector of three control variables, $\phi$ the corresponding vector of parameters; whereas $\eta_{t}$ is the time effect and $\epsilon_{it}$ is the error term. The three control variables that are included are the $\log$ of the initial GDP (PPP) per capita, the increase in natural resource exports and the initial \textit{complexity outlook index} (COI) of the country. The first one helps us explain the catch-up effect, which states that the developing countries have potential to grow at faster rates, when compared to the developed ones. Since economic complexity does not capture the effects of exporting natural resources, we include the increase in natural resource exports in constant dollars as a share of initial GDP (natural resources match SITC categories 0, 1, 2, 3, 4 and 68, same as Sachs and Warner \cite{Sachs-1995}) as the second control variable. In addition, in every estimation, $c_{i,t}$ is accompanied by the COI index that measures how many different products are near a country's set of productive capabilities, according to the capability-driven model for economic complexity. The role of this index, according to \cite{Hidalgo-2014}, stems by the fact that "the countries with a low complexity outlook have few nearby products and will find it difficult to acquire new capabilities and increase their economic complexity".

Table \ref{Table:growthregECI} provides the results from the performed regressions for the effect that the country complexity measure obtained through the MR method has on the economic growth. The first column shows the estimated parameters when growth is regressed only on the first two control variables, while the second adds the aggregated ECI and COI to the regression. From the regression it can be easily concluded that the aggregated ECI is a significant predictor of long term economic growth at any level. The third column replaces the aggregated ECI with the disaggregated ECI. As a consequence, the value of $R^2$ increases almost by 10 percentage points. In order to test which model would be preferred based on the data, in the fourth column we combine the two indices (See Brooks \cite{Brooks-2014} p. 159 for the procedure). In this model the aggregated ECI is no longer significant, implying that the model with the disaggregated ECI is preferred. This is no coincidence since the aggregated ECI is constructed by distinguishing fewer number of goods, and thus it contains less information about the productive capabilities embedded in them. In the last two columns we compare the performance of the aggregated ECI and the simplest measure for productive capabilities, the aggregated diversity. Column $\rom{5}$ suggests that the aggregated diversity significantly explains the changes in growth only if the critical level is above 10 percent, while column $\rom{6}$ states that the aggregated ECI is a better representative of the data. More precisely, when the aggregated ECI and the aggregated diversity are combined in one regression, the coefficient of aggregated diversity becomes negative and loses significance, whereas the significance of the aggregated ECI remains. This conclusion is justified even though the colinearity between these variables is very high, and it is well known that if two highly colinear variables are included as explanatory variables in a regression, it may happen that none of them remain significant (we measure the level of colinearity as the Pearson correlation controlled for the log of the initial GDP per capita, natural resource exports and COI, which is 0.66).

\begin{table} [b!]
\begin{adjustwidth}{-0.5in}{0.5in}
\caption{\textbf{Aggregated Economic Complexity Index and Growth.}} 
\centering
\setlength{\tabcolsep}{0.6cm}
\begin{tabular}{|c|d{3}|d{3}|d{3}|d{3}|d{3}|d{3}|d{3}|} 
\hline
\multicolumn{7}{|c|}{\textbf{Dependent Variable: Growth in GDP pc}} \\
\multicolumn{7}{|c|}{\textbf{1988-1998, 1998-2008}} \\
\hline
\multicolumn{1}{|c|}{\textbf{Variable}} &\multicolumn{1}{|c|}{(\rom{1})} &\multicolumn{1}{|c|}{(\rom{2})} &\multicolumn{1}{|c|}{(\rom{3})} &\multicolumn{1}{|c|}{(\rom{4})} &\multicolumn{1}{|c|}{(\rom{5})} &\multicolumn{1}{|c|}{(\rom{6})} \\
\hline
\multicolumn{1}{|l|}{Income per capita, logs} & -0.027^{**} & -0.041^{***}& -0.106^{***} & -0.106^{***}& -0.033^{**}& -0.041^{***}\\\hline
& (0.013) & (0.015) & (0.019) & (0.019) & (0.014) & (0.015) \\\hline
\multicolumn{1}{|l|}{Increase in NR exports} & 0.311^{*} & 0.325^{*}& 0.311^{**}& 0.306^{**}& 0.304^{*}& 0.326^{*}\\\hline
& (0.166) & (0.166) & (0.132) & (0.133)& (0.165) & (0.160) \\\hline
\multicolumn{1}{|l|}{Initial aggregated ECI }  && 0.042^{***} &&-0.008& &  0.045^{**}\\\hline
&& (0.015) && (0.016)& & (0.022) \\\hline
\multicolumn{1}{|l|}{Initial COI} && 0.005^{**} & 0.005^{***}&0.005^{***}&0.004^{**}&0.005^{**} \\\hline
&& (0.002) &(0.002)&(0.002)& (0.002) &(0.002)\\\hline
\multicolumn{1}{|l|}{Initial ECI}&&& 0.118^{***}& 0.123^{***}& &\\\hline
&&& (0.023) & (0.026)& &\\\hline
\multicolumn{1}{|l|}{Initial aggregated Diversity} &&&&&0.010^{*}& -0.002 \\\hline
&&&&& (0.006)   & (0.009) \\\hline
\multicolumn{1}{|l|}{Constant} & 0.468^{***} & 0.631^{***} & 1.187^{***}&1.189^{***} & 0.511^{***}& 0.637^{***}\\\hline
& (0.117) & (0.134) & (0.175) & (0.175)& (0.117)  & (0.139) \\\hline
\multicolumn{1}{|l|}{Observations} &\multicolumn{1}{|c|}{210} &\multicolumn{1}{|c|}{210} &\multicolumn{1}{|c|}{210} &\multicolumn{1}{|c|}{210} &\multicolumn{1}{|c|}{210}  &\multicolumn{1}{|c|}{210} \\\hline
\multicolumn{1}{|l|}{$R^{2}$} & 0.202 & 0.239 & 0.334 & 0.334& 0.224 &  0.239\\\hline
\multicolumn{1}{|l|}{Year FE} &\multicolumn{1}{|c|}{Yes} &\multicolumn{1}{|c|}{Yes} &\multicolumn{1}{|c|}{Yes} &\multicolumn{1}{|c|}{Yes} &\multicolumn{1}{|c|}{Yes} &\multicolumn{1}{|c|}{Yes} \\

\hline
\end{tabular}
\begin{flushleft}
 Notes: Liberia in both periods and Turkmenistan in 1998 were excluded since they were extreme outliers. Standard errors clustered by cross-section shown in parentheses. ***$p<0.01$, **$p<0.05$, *$p<0.1$

\end{flushleft}

\label{Table:growthregECI}
\end{adjustwidth}
\end{table}

\paragraph*{FCM Regressions:}
In the original works which introduced the fitness-complexity method\cite{Tacchella-2012, Cristelli-2013}, the authors use the Harmonized System 2007 nomenclature for goods in order to estimate the (disaggregated) intensive fitness (IF), which is slightly different from the one used in \cite{Hidalgo-2009, Hidalgo-2014} and in our work. Moreover, as we noted in the \nameref{Materials and Methods} section, the FCM in its original version may result in complexity indices converging to zero values,  thus failing to display any significant results in typical regression analysis. In order to deal with this issue, and with the aim to produce a fair comparison when different goods nomenclature is used, we re-estimate the disaggregated IF with the modified fitness-complexity method (M-FCM), by relying on the SITC rev.2 4-digit classification. Then, in Table \ref{Table:growthregIF} we redo the growth regressions with the logs of the estimated aggregated IF. Afterwards, we compare its performance with the logs of the disaggregated IF and the aggregated diversity. The results are in line to the ones obtained with the MR method: first, the $\log$ of the aggregated IF is a significant explanatory variable of growth at any level; second, the econometric model which includes the aggregated IF performs worse than the disaggregated IF. The only difference appears in the model in which we compare the performance of the aggregated IF and the aggregated diversity, where now both variables are significant (although the aggregated diversity is significant only at 10$\%$). However, as in the case of the comparison between the aggregated ECI and the aggregated diversity, in this combined model the relationship between aggregated diversity and growth becomes negative. This observation appears because the collinearity between the aggregated IF and aggregated diversity is also very high (they have a Pearson correlation of 0.89 when controlling for for the log of the initial GDP per capita, natural resource exports and COI). Altogether, it implies that the aggregated IF is a more robust variable than aggregated diversity, when it comes to the explanation of the long term changes in income. This allows us to conclude that the prediction model based on the $\log$ of aggregated IF performs better than the model based on aggregated diversity.

\begin{table} [b!]
\begin{adjustwidth}{-0.5in}{0.5in}
\caption{\textbf{Aggregated Intensive Fitness and Growth.}} 
\centering
\setlength{\tabcolsep}{0.6cm}
\begin{tabular}{|c|d{3}|d{3}|d{3}|d{3}|d{3}|d{3}|d{3}|} 
\hline
\multicolumn{7}{|c|}{\textbf{Dependent Variable: Growth in GDP pc}} \\
\multicolumn{7}{|c|}{\textbf{1988-1998, 1998-2008}} \\
\hline
\multicolumn{1}{|c|}{\textbf{Variable}} &\multicolumn{1}{|c|}{(\rom{1})} &\multicolumn{1}{|c|}{(\rom{2})} &\multicolumn{1}{|c|}{(\rom{3})} &\multicolumn{1}{|c|}{(\rom{4})} &\multicolumn{1}{|c|}{(\rom{5})} &\multicolumn{1}{|c|}{(\rom{6})} \\
\hline
\multicolumn{1}{|l|}{Income per capita, logs} & -0.027^{**} & -0.045^{***}& -0.068^{***}& -0.067^{***}& -0.033^{**}& -0.051^{***}\\\hline
& (0.013) & (0.015) & (0.017) & (0.017)& (0.014)  & (0.015) \\\hline
\multicolumn{1}{|l|}{Increase in NR exports} & 0.311^{*} & 0.312^{*}& 0.321^{**}&0.319^{**}& 0.304^{*}&  0.311^{*}\\\hline
& (0.166) & (0.162) & (0.155) & (0.156)& (0.165)  & (0.160) \\\hline
\multicolumn{1}{|l|}{Initial aggregated IF,logs}  && 0.046^{***} && -0.014&& 0.107^{**}\\\hline
&& (0.017) && (0.021)& & (0.042) \\\hline
\multicolumn{1}{|l|}{Initial COI} && 0.004^{**} & 0.005^{***}&0.005^{***} &0.004^{**}&0.005^{**} \\\hline
&& (0.002) &(0.002)&(0.002)& (0.002) &(0.002)\\\hline
\multicolumn{1}{|l|}{Initial IF, logs}&&& 0.086^{***}& 0.095^{***}& &\\\hline
&&& (0.023)& (0.029) &  &\\\hline
\multicolumn{1}{|l|}{Initial aggregated Diversity} &&&& &0.010^{*}& -0.026^{*} \\\hline
&&&&& (0.006)   &(0.015) \\\hline
\multicolumn{1}{|l|}{Constant} & 0.468^{***} & 0.692^{***} & 0.900^{***} & 0.884^{***}&0.511^{***}& 0.888^{***}\\\hline
& (0.117) & (0.137) & (0.158) & (0.151)& (0.117)  & (0.181) \\\hline
\multicolumn{1}{|l|}{Observations} &\multicolumn{1}{|c|}{210} &\multicolumn{1}{|c|}{210} &\multicolumn{1}{|c|}{210} &\multicolumn{1}{|c|}{210} &\multicolumn{1}{|c|}{210}  &\multicolumn{1}{|c|}{210} \\\hline
\multicolumn{1}{|l|}{$R^{2}$} & 0.202 & 0.241 & 0.296 & 0.297 & 0.224 & 0.254\\\hline
\multicolumn{1}{|l|}{Year FE} &\multicolumn{1}{|c|}{Yes} &\multicolumn{1}{|c|}{Yes} &\multicolumn{1}{|c|}{Yes} &\multicolumn{1}{|c|}{Yes} &\multicolumn{1}{|c|}{Yes} &\multicolumn{1}{|c|}{Yes} \\

\hline
\end{tabular}
\begin{flushleft}
	Notes: For consistency with the aggregated ECI regressions, Liberia in both periods and Malawi in 1998 were excluded from the sample. Standard errors clustered by cross-section shown in parentheses. ***$p<0.01$, **$p<0.05$, *$p<0.1$

\end{flushleft}

\label{Table:growthregIF}
\end{adjustwidth}
\end{table}
	
To make a more adequate presentation of the marginal effects of both aggregated ECI and the $\log$ of the aggregated IF, we standardize their estimated coefficients by multiplying them by the ratio of the standard deviations of the independent and dependent variables \cite{Herzer-2012}. The standardized coefficients imply that, a one standard deviation increase in the aggregated ECI is associated with an increase in the growth variable equal to 18.6$\%$ of a standard deviation in that variable, while a one standard deviation increase in the $\log$ of aggregated IF increases the same growth variable by 20.4$\%$ of its standard deviation. On the other hand, the disaggregated ECI and the $\log$ of the disaggregated IF estimate that the marginal effect of country complexity over growth is, respectively, 55.7 percent and 37.3 percent. From this, it can be said that the aggregated complexity measures have similar effect on economic growth, and, in addition with the nearly identical $R^{2}$, it appears that the aggregation and the inclusion of services lessen the discrepancy between the marginal effect and the explanatory power of the two metrics. More importantly, the magnitude of the standardized coefficients is only slightly smaller than the increase in natural resource exports (which is between $25\%$ and $27\%$ of a standard deviation in both regressions), thus allowing us to conclude that an increase in the aggregated economic complexity has an economically large effect.

While the performed regressions are appropriate for capturing the explanatory potential of the complexity indices, they are not appropriate for growth prediction due to several reasons. First, we recall that the $R^2$ is a standardized measure, bounded between 0 and 1, which estimates how much of the variance of the dependent variable can be explained by the model. If the regression is supposed to be used for predictions, it should have an $R^2$ very close to $1$. However, in each of the performed regressions this measure is fairly low (the best performer is the disaggregated ECI with an estimate of $0.33$). In fact, as Table \ref{Table:growthregECI} and Table \ref{Table:growthregIF} show, aggregated country complexity and aggregated diversity have similar prediction power, which is an additional argument against the regression analysis. As discussed in \cite{Tacchella-2012, Cristelli-2013, Cristelli-2015},  this type of regression analysis fails to spot the heterogeneous features that change over time and between countries. We conclude that the use of the complexity measures for prediction requires more sophisticated methods (see \cite{Cristelli-2015} for related discussion). 

\subsection*{Services vs. Goods}
\begin{figure}[t!]
\begin{adjustwidth}{0in}{0in}
\includegraphics[width=16cm]{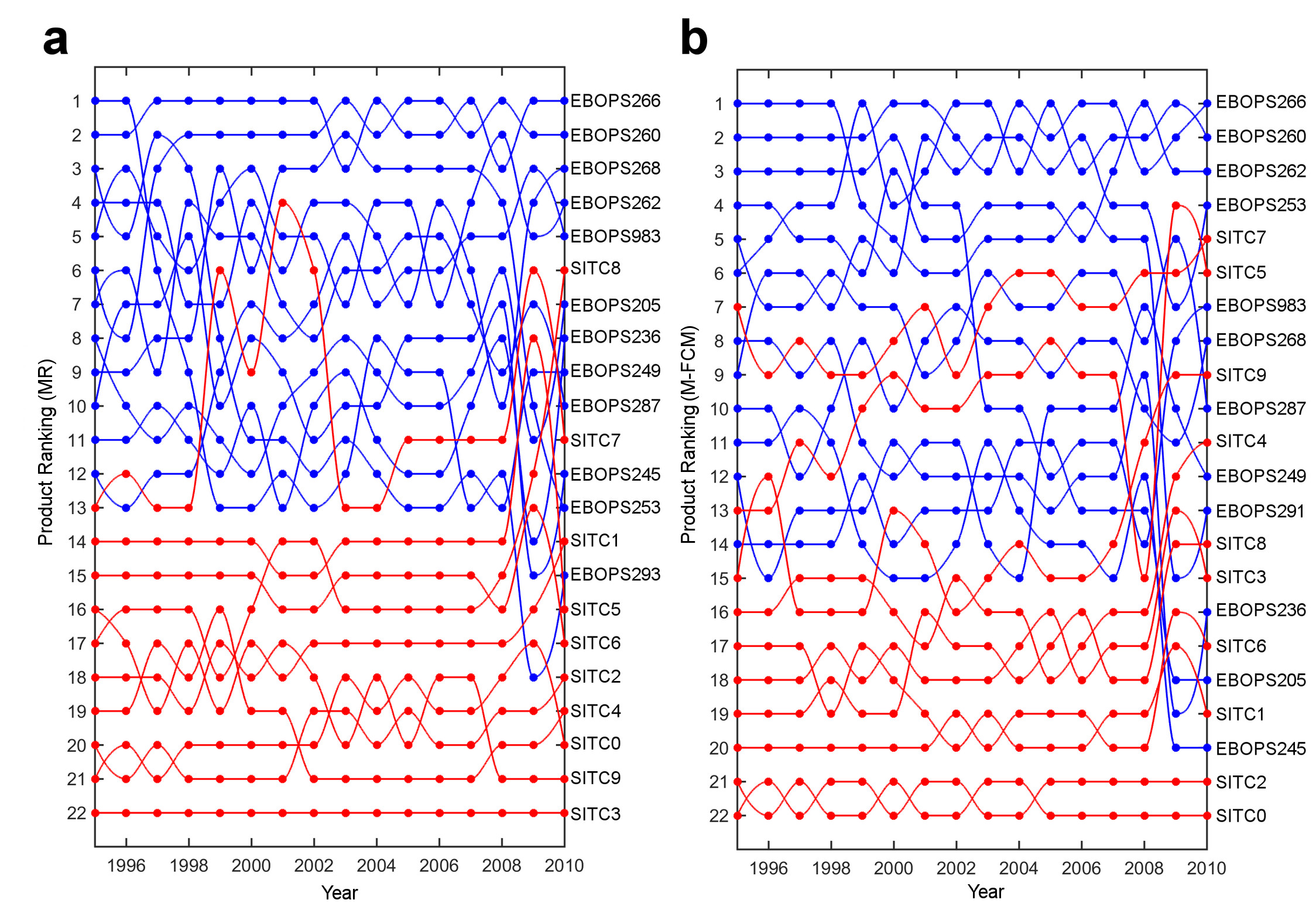}
\caption{\textbf{ Evolution of services and goods complexity (1995-2010).} \textbf{a)} Product ranking estimated through MR. \textbf{b)}Same as a) only estimated through M-FCM. \textbf{a-b} Services: blue, goods: red.}  \label{psplot}
\end{adjustwidth}
\end{figure}

Here we investigate the dynamics of the aggregated Product Complexity Index estimated through both MR and M-FCM. We begin our analysis by presenting a visualization of the aggregated product complexity ranking in Fig. \ref{psplot}. It can be easily noticed that most services have higher complexity than goods. This result can be statistically supported by regressing the aggregated PCI against a dummy variable that captures the effect of services and fixing the time effects:
\begin{align*}
p_{jt}=\mu_g + \Delta \mu_s D_s + \eta_t +\epsilon_{jt}
\end{align*}
where $p_{jt}$ is either the linear (estimated through MR) or the nonlinear (estimated through M-FCM) product complexity index for good or service $j$ at time $t$, $\mu_g$ is the average good complexity, $\Delta \mu_s$ is an estimate of the additional average service complexity; $D_s$ is the dummy variable; $\eta_t$ the time effects and $\epsilon_{jt}$ is the error component. 

The results of the regressions are provided in Table \ref{Table:pscireg}. From column $\rom{1}$ it can be concluded that for the method of reflections, after controlling for the year, the average service complexity is higher than the average good complexity by $1.59$ standard deviations. The modified fitness-complexity method produces similar results, as presented in column $\rom{2}$ – the average service complexity is $2.77$ times higher than the average good complexity.

\begin{table} [b!]
\begin{adjustwidth}{0in}{0in}
\caption{\textbf{PCI Regressions.}} 
\centering
\setlength{\tabcolsep}{1.5cm}
\begin{tabular}{|c|d{3}|d{3}| } 
\hline
\multicolumn{3}{|c|}{\textbf{Dependent Variable: Product Complexity Index}} \\
\multicolumn{3}{|c|}{\textbf{1995-2010}} \\
\hline
\multicolumn{1}{|c|}{\textbf{Variable}} &\multicolumn{1}{|c|}{(\rom{1})} &\multicolumn{1}{|c|}{(\rom{2})} \\
\hline
\multicolumn{1}{|l|}{Service dummy} & 1.594^{***} & 0.901^{***}\\\hline
& (0.225) & (0.233) \\\hline
\multicolumn{1}{|l|}{Constant} & -0.870^{***} & 0.508^{***}\\\hline
& (0.201) & (0.009) \\\hline
\multicolumn{1}{|l|}{Observations} & \multicolumn{1}{|c|}{352} & \multicolumn{1}{|c|}{352} \\\hline
\multicolumn{1}{|l|}{$R^{2}$} & 0.660 & 0.268\\\hline
\multicolumn{1}{|l|}{Year FE} &\multicolumn{1}{|c|}{Yes} &\multicolumn{1}{|c|}{Yes} \\

\hline
\end{tabular}
\begin{flushleft}
	Note: Standard errors clustered by cross-section shown in parentheses. ***$p<0.01$ 
\end{flushleft}

\label{Table:pscireg}
\end{adjustwidth}
\end{table}

In the spirit of a capability-driven productive structure, it seems that services require more capabilities for their delivery. But why is this the case?

As Hausmann and Hidalgo \cite{Hidalgo-2011} argue, the production of a certain good requires the country to have the necessary capabilities associated with it. In other words, a link between a country and a product (be it a good or a service) exists if the country has the capabilities required to produce it. The framework which relates products and capabilities may be used to qualitatively describe complexity of services. In particular, by looking at the bottom of the product complexity rankings we find that according to MR, SITC 3: Mineral Fuels, Lubricants and related materials is the least complex product in each year, while according to the M-FCM the least complex products are SITC 0: Food and live animals and SITC 2: Crude materials, inedible, except fuels. Disaggregating them, we can get various raw materials whose production does not rely as much on the presence of capabilities, as it does on the geographical luck of the country.

By moving further up in the rankings we observe that in the majority of the examination period SITC 7: Machinery and transport equipment represents the most complex aggregated good. For a country to be able to produce machineries such as combustion engines or x-ray equipment, besides using natural resources as inputs, it has to combine other capabilities, for example the capabilities embedded in its workers. The positioning of the linear aggregated products is well reflected in the SITC rev.2 4-digit ranking provided in the Observatory of Economic Complexity \cite{Simoes-2010}.  

At the very top we can find services such as EBOPS 266: Royalties and Licence fees. We believe that, in the sense of \cite{Hidalgo-2011}, the production of these services requires the combination of the complex knowledge embedded in the goods and in the other services that a country produces. In addition, we argue that as a country's product space becomes saturated with goods, the contribution of economic complexity to future economic growth begins to decrease, which is in line with the observations in \cite{Hidalgo-2014}. In this context, our results suggest that probably the easiest way for maintaining the effect of the evolution of the productive structure is by introducing services whose set of required capabilities is similar, but a little more complex; for instance, developing intangible assets such as patents requires the presence of sophisticated machineries and an additional input of capabilities. It is worth noting that our argument is in accordance with the conclusions in \cite{Saez-2014} where it was indicated that services tend to grow in importance for the economy as the level of a country's development rises.

\subsection*{Aggregated vs. Disaggregated Metrics}
\begin{figure}[t!]
\begin{adjustwidth}{0in}{0in}
\includegraphics[width=18cm]{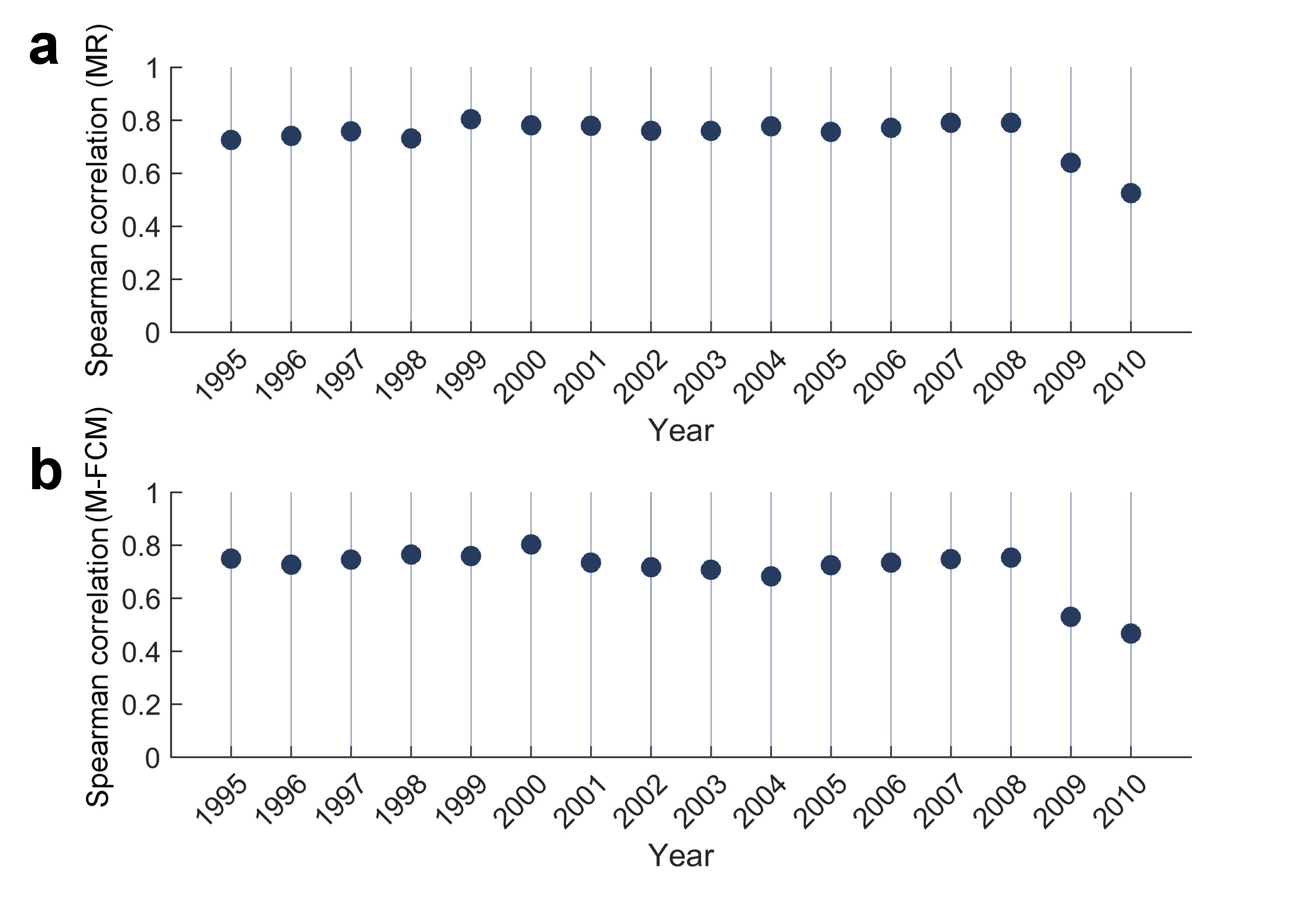}
\caption{\textbf{Correlation between the aggregated and disaggregated metrics}. \textbf{a)} Yearly Spearman correlation between the aggregated and disaggregated estimated through MR. \textbf{b)} same as  \textbf{a)} for the aggregated and disaggregated estimated through M-FCM. \label{fig:aggdis}}
\end{adjustwidth}
\end{figure}

On the on hand, the ability of the aggregated complexity measures (with goods and services included in the model) to significantly explain the long term growth of the countries implies that these measures are somewhat similar to the disaggregated complexity measures (obtained with goods only). Yet, the conclusion from the previous section that services are, in general, more complex than goods indicates that the aggregated metrics provide different information about the present competitiveness and the future potential of the countries. In order to get a better presentation of the (dis)similarity of the two type of metrics and to assess the effects of goods aggregation and inclusion of services on the productive structure of the countries, in Fig. \ref{fig:aggdis} we show the yearly Spearman Rank correlation between the country rankings based on aggregated (with goods and services) and disaggregated (goods) data. It can be observed that there is a strong correlation between the aggregated and disaggregated metrics until 2008 (Only in 2004 the correlation between the aggregated and disaggregated IF is under 0.7). While this similarity can be used as an evidence that the aggregated metrics are a good approximation of the productive structure (if we take the disaggregated metrics as benchmark), the small distinction reveals the different information which they produce. These differences, we believe, can be attributed more to the inclusion of services in the model than to the aggregation of goods. Our reasoning is founded on the conclusion from the previous section that services are, in general, more complex than goods and that the ranking of aggregated goods complexity exhibits nearly the same structure as the disaggregated goods complexity ranking \cite{Simoes-2010}, which means that in the aggregated rankings service-oriented economies become more complex at the expense of the goods-oriented economies. This conclusion is additionally supported by the procedure of the two methods - both of them, due to the initial aggregated diversity and ubiquity profiles, in an iterative manner discover even bigger differences in the complexity score of service-oriented and goods-oriented countries (this characteristic is more pronounced in M-FCM).

In the aftermath of the Global Crisis in 2008, we observe a decrease in the correlation between the aggregated and the disaggregated indices (both for ECI and IF), dropping under $0.65$ in 2009 and under $0.55$ in 2010. This suggests that the Financial crisis and the following recession (accompanied by the European Sovereign-Debt crisis), had a more profound impact on the productive structure of the world economies. It is our belief that this impact can be attributed to the decreased volume in trade \cite{IMF-2009} that was more pronounced in goods than in services \cite{Levchenko-2009}. As a result, the complexity of some of the the aggregated goods increased (as shown in Fig. \ref{psplot}). To explain this effect, we need to recall the iterative relation between complexity of countries and products. The main argument comes from the fact that the goods exports were not affected in the same way for all countries: some of the countries managed to pertain same or similar level of goods export; others, however, exhibited sharper decrease in their exports. Correspondingly, some of the goods that were produced by the majority of the countries before the crisis, were now produced by less countries. As a consequence, the complexity of the affected goods increased, yielding an effective increase of the complexity of their major  producers. This process eventually led to an increase of the complexity of some of the countries (those less affected by the crisis), at the expense of the decrease in the complexity of other countries (those most affected more by the crisis in the global goods trade).

These changes were additionally fueled by the nature of some of the service types that we classified. In particular, we observe that in light of financial crises, the financial services (which are usually the most complex services according to our model) are subject to a major decrease in trade, thus significantly affecting complexity indices and country rankings. It is also possible that the austerity measures imposed in some countries with the aim to handle the Sovereign-Debt crisis, affect service exports in a different way that they affect goods exports \cite{WTO-2011}. The validation of this argument, however, requires a deeper analysis which is out of the scope of this work. In any case, services such as finance, information and royalties still remained the most complex products, and hence the countries which maintained a comparative advantage in these products, the most competitive ones. As a matter of fact, as we will see later on, the same countries were able to recover faster from the recession and to display significant progress after it. We discuss this argument in more detail in the following section. 

\subsection*{Country Specifics}

\begin{figure}[t!]
\begin{adjustwidth}{0in}{0in}
\includegraphics[width=17cm]{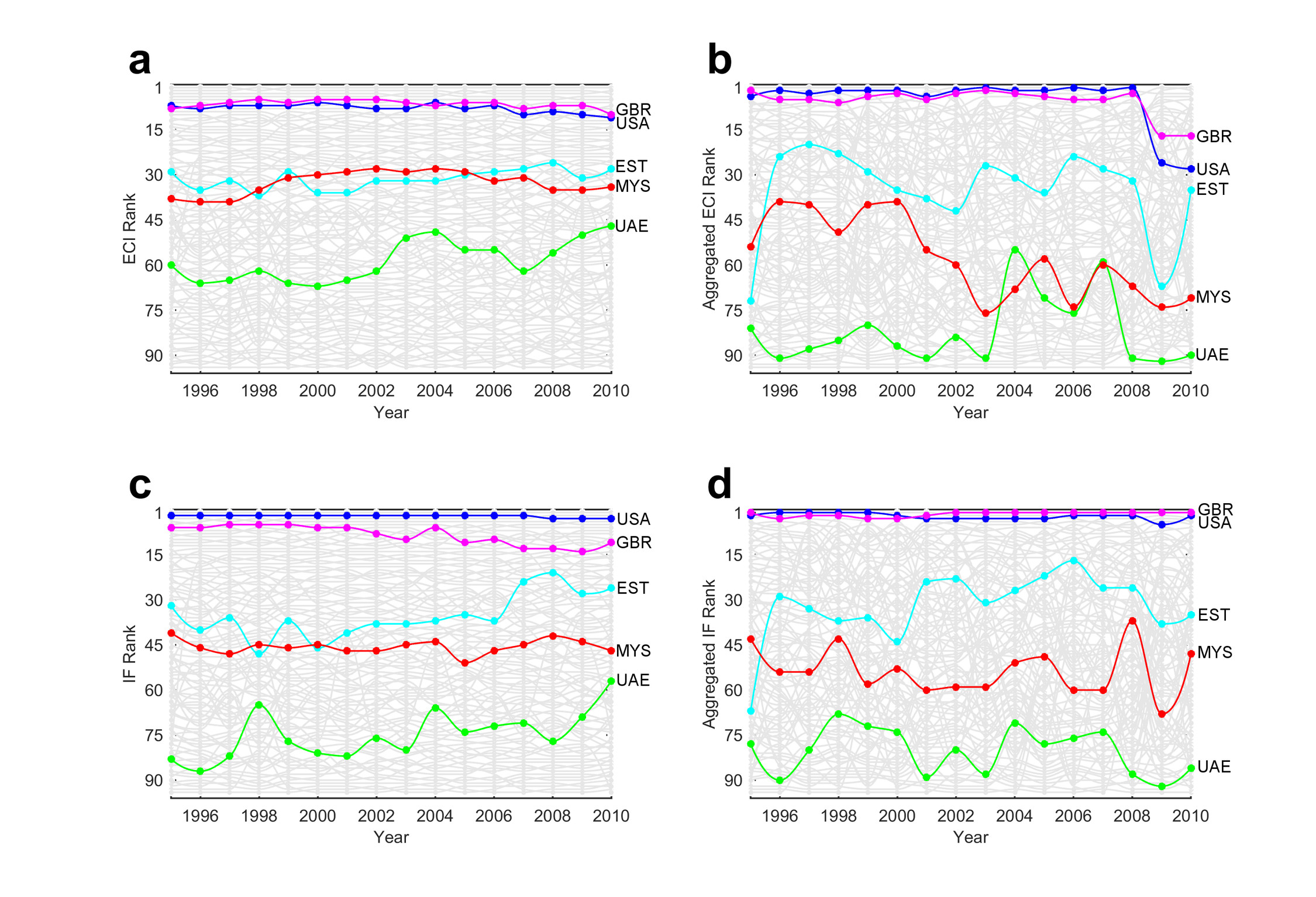}
\caption{\textbf{Country Dynamics (1995-2010).} \textbf{a)} ECI rankings movement of the countries. We thereby highlight China, Estonia, Ethiopia, Malaysia, Switzerland, the United Arab Emirates, United Kingdom and the United States.  \textbf{b)} Same for aggregated ECI. \textbf{c)} IF rankings movement of the countries. \textbf{d)} Same for aggregated IF. \textbf{a-d} Three digit iso codes are used as country abbreviations. \label{fig:unweighted}}
\end{adjustwidth}
\end{figure}

\paragraph*{Country Dynamics:}
As a means to provide a more detailed analysis of the country rankings, in Fig. \ref{fig:unweighted} we display the dynamics of the aggregated and disaggregated ECI and IF. Thereby, we highlight several countries (Estonia, Malaysia, United Arab Emirates (UAE), United Kingdom (UK) and the United States (USA)) with different economic attributes. Here countries are ranked according to their ECI (Fig. \ref{fig:unweighted}a), aggregated ECI (Fig. \ref{fig:unweighted}b, IF (Fig. \ref{fig:unweighted}c) and aggregated IF (Fig. \ref{fig:unweighted}d) evaluated between 1995 and 2010. Due to limited data, the ranking includes only 94 countries.

It can be noticed that the country complexity rankings produced through aggregated data are more volatile than the ones produced taking account disaggregated data. We believe that both the aggregation of the data and the nature of the services are behind this observation. The argumentation goes along the observation that, although complexity indices do not account for the volume of trade, they still account for its presence. We recall that the aggregated ECI and IF rely on aggregated data which effectively decreases the number of products and, by that, the dimension of the adjacency country-product matrix. With this, the disappearance of a link or the occurrence of new link in the country-product matrix due to data variations  has a far bigger impact on the complexity indices than in the case with disaggregated goods data.

The years after 2008 are characterized by an even larger volatility in the country rankings which, we believe, cannot be attributed solely to the aggregation of data. Based on these results we argue that the aggregated measures are most probably more susceptible to economic crises (as compared to the disaggregated ones), resulting in more frequent but also bigger changes in the rankings.

\paragraph*{Countries with developed service sector:}
As observed from the presented results, the inclusion of services affects significantly the complexity indices and the rankings of individual countries. Specifically, the fact that the services appear to be on average more complex than goods (i.e. rank higher on the product complexity scale), positions (in general) countries with developed service sector higher in the economic complexity ranking. This holds in particular for countries such as USA and UK. According to Fig. \ref{fig:unweighted}, until 2008 UK is ranked higher in both aggregated rankings (aggregated ECI/aggregated IF), as compared to the disaggregated (goods only) rankings. After 2008 we observe a drop in its aggregated ECI rankings which, interestingly, is not reflected in the aggregated IF rankings.

\begin{figure}[t!]
\begin{adjustwidth}{0in}{0in}
\includegraphics[width=16cm]{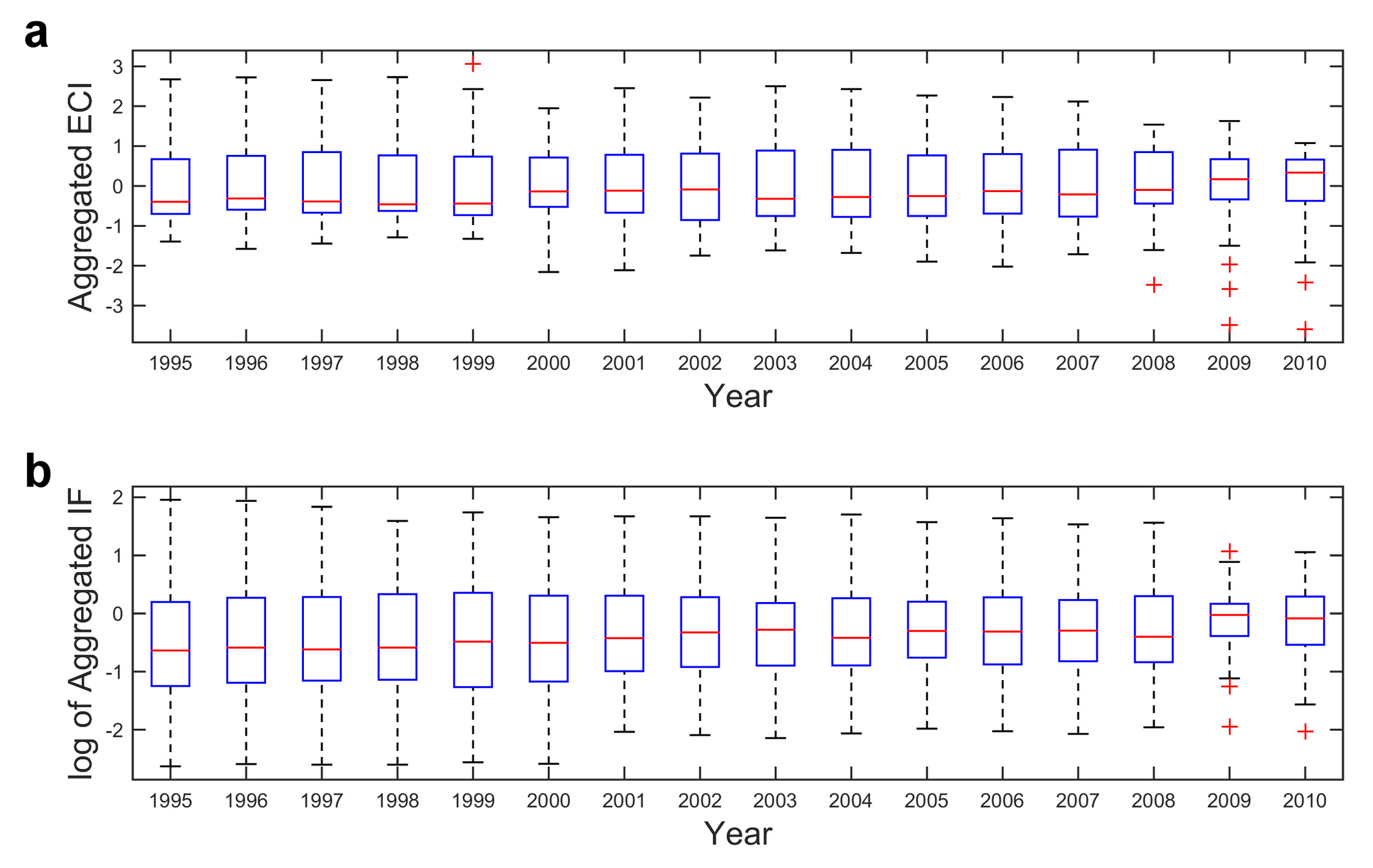}
\caption{\textbf{Changes in the distribution of complexity.} \textbf{a)} Box plots for the aggregated ECI (1995-2010)  \textbf{b)} Same for aggregated IF. \textbf{a-b} In each year, red horizontal lines indicate the median, whereas the blue horizontal lines are, respectively, the first and third quartile of the distribution. Red crosses indicate outliers. They are identified as being observations which are, either greater than the sum of the third quartile and 3/2 of the interquantile range or less than the difference between the first quartile and 3/2 of the interquantile range. These thresholds are displayed with black horizontal lines. \label{fig:boxplot}}
\end{adjustwidth}
\end{figure}

We offer the following heuristic explanation for this discrepancy. First, we observe that in the aftermath of the financial crisis there are substantial changes in the structure of the $M$ matrix, which now may be described as being "more uniform". This is to say that, as a result of the crisis, there is reduction in the differences between the countries' export baskets. This feature can be deduced from Fig. \ref{fig:boxplot} where we display the yearly Box plots for the aggregated ECI and the log of aggregated IF. Specifically, we observe an increase in the median complexity and a significant decrease in the interquartile range in the yearly distributions of both complexity metrics. These changes in the distribution, are more pronounced in the MR model, due to the linear relation between the complexity of a product and the complexity of its exporters. Consequently, there is a decrease of the complexity indices (ECI) for the countries whose product portfolio includes sophisticated services, which is eventually reflected in the complexity rankings. In the FCM model, on the other hand, the complexity of a product exhibits a nonlinear relation with the complexity (i.e. fitness) of its exporters, being mostly dominated by the fitness of the least complex exporter. The key point is that, with the structure of $M$ now being more uniform, the variations of the complexity (fitness) indices of the countries decreases. However, on average, the change in the fitness of the least complex exporters is less pronounced than the change of the fitness of the most complex exporters. In other words, it seems that, compared to MR, the FCM dynamics is less affected by short term economic fluctuations.

During the crisis, UK entered a recession, and experienced the second highest downturn among the G7 countries (G7 is a group consisting of the most developed countries in the world). Following another stagnation during 2013, UK became the fastest growing developed economy \cite{Telegraph}. Similar fluctuations, to those of UK, manifested USA in both aggregated and disaggregated rankings (although in the aggregated rankings USA is ranked slightly lower than UK since it is not as reliant on production of services). Indeed, USA's economy also showed nearly the same movements - it fell under a deep recession in 2008 and recovered relatively fast. The differences between the complexity dynamics based on aggregated goods and services and the complexity dynamics based only on (disaggregated) goods of UK and USA are even more pronounced when we compare our aggregated complexity metrics (which account for both goods and services) to the ones calculated by only using the SITC Rev.2 1-digit aggregated goods. Therefore, we argue that by including the services, the aggregated metrics present a more realistic picture about the long term potential displayed by these two countries. 

The distinction between the rankings is even more pronounced when examining countries like Estonia and Malaysia which have similar wealth and disaggregated rankings through the years, but different GDP structure. Estonia's economy relies heavily on services, whereas Malaysia's is almost equally reliant on both goods and services. As a consequence, Estonia is usually ranked far higher in the aggregated rankings, and has experienced bigger rises and falls over the years. For comparison, at the same time Estonia averaged a bigger yearly GDP per capita growth ($4.99\%$, whilst Malaysia had $2.65\%$), had bigger standard deviation in it ($7.07$ to $4.34$) and was eventually classified as advanced economy by IMF \cite{IMF-2011}.

A separate type of countries are the natural resource exporters. According to Hausmann and Hidalgo \cite{Hidalgo-2014}, some of these countries witnessed tremendous development over the years not because of their complex economy, but because of the abundance of raw materials who do not require many capabilities to be produced. An example of such a country is the United Arab Emirates, which is ranked relatively low in the rankings. The aggregated ECI and IF rank UAE even lower in some years as the revealed comparative advantage that UAE had in some complex products is lost by the aggregation of the data. 

In agreement with all these observations, we argue that the aggregated metrics (with services included) can be used as a better representation of the productive structure of some of the developed economies. We base our argument on the fact that the examined countries (particularly USA and UK) were among the first to experience the post-industrial transition from a manufacturing-based economy to a service-based economy during the last few decades of the twentieth century \cite{Bell-1974}. This transition, characterized with a declining manufacturing sector, yields an effective decrease of the complexity of a country, when estimated only by using goods data.
 
While the same conclusions hold for some of the other developed countries, as in the case of Switzerland for example, they appear to not be true for Germany, Japan and Italy. In fact, their rankings are reversed: in the rankings based solely on goods they are among the most complex economies, whereas they are placed on average lower in the rankings where services are included. We believe that this discrepancy is due to the difference in the speed of transition. It is known that over the years, Germany, Japan and Italy had the lowest increase in service employment among the developed countries \cite{Castells-2011}, which implies that they preserved the diversity in the goods sector and did not develop as diversified service sector as other developed countries.

On one side, it may be the case that economic complexity indices based only on goods are slightly biased towards these type of economies. This could especially be true for Italy, a country which exhibits great discrepancy in the rankings: in the aggregated country rankings it is always placed outside the top ten, meaning that it definitely does not have as competitive service sector as other economies, and yet, according to the disaggregated IF, is constantly among the five most complex countries. Actually, Italy's economic system was hit hard during the Great Recession and, consequently, this led to the nations Sovereign-Debt crisis. While this fall in the long run potential can be predicted from the country's movements in the aggregated IF rankings in the late 2000s (See the next paragraph), it can not be, in any way, sensed from the disaggregated dynamics (based solely on disaggregated goods data). 

However, on the other side, it may also be that aggregated rankings which include services are biased towards the economies with sophisticated service sector, as it is the example with the USA and, in particular, with UK. More precise insights would definitely be obtained by performing an analysis based on finely disaggregated goods and service data, something which is not possible due to current service nomenclature and data availability.
   
\paragraph*{The PIIGS nations:}
PIIGS is a derogatory term which refers to the nations of: Portugal, Ireland, Italy, Greece and Spain; five European Union member states that were unable to refinance their government debt or to bail out over-indebted banks on their own during the Sovereign-Debt crisis of 2009. Their downturn was discussed from an economic complexity perspective in \cite{Cristelli-2013}. According to the authors, the fragility of the PIIGS is mostly of financial origin and is less related to their productive structure. They strengthen their argument by pointing out that there are different regimes in the fitness-income plane, and that the PIIGS are in the one where the developed nations are stationed. The economies in this regime are less reliant on their productive structure for future growth since they already have a diversified goods export basket; i.e. they have saturated their product space. 

While we mostly agree with this argument, we would also like to investigate the possibility that the fragility of the PIIGS countries may be, to a certain extent, attributed to the structure of their service sector. To obtain some further clues, in the following we concentrate on the economic complexity of these countries where services are added in the model.

\begin{figure}[b!]
\begin{adjustwidth}{0in}{0in}
\includegraphics[width=17cm]{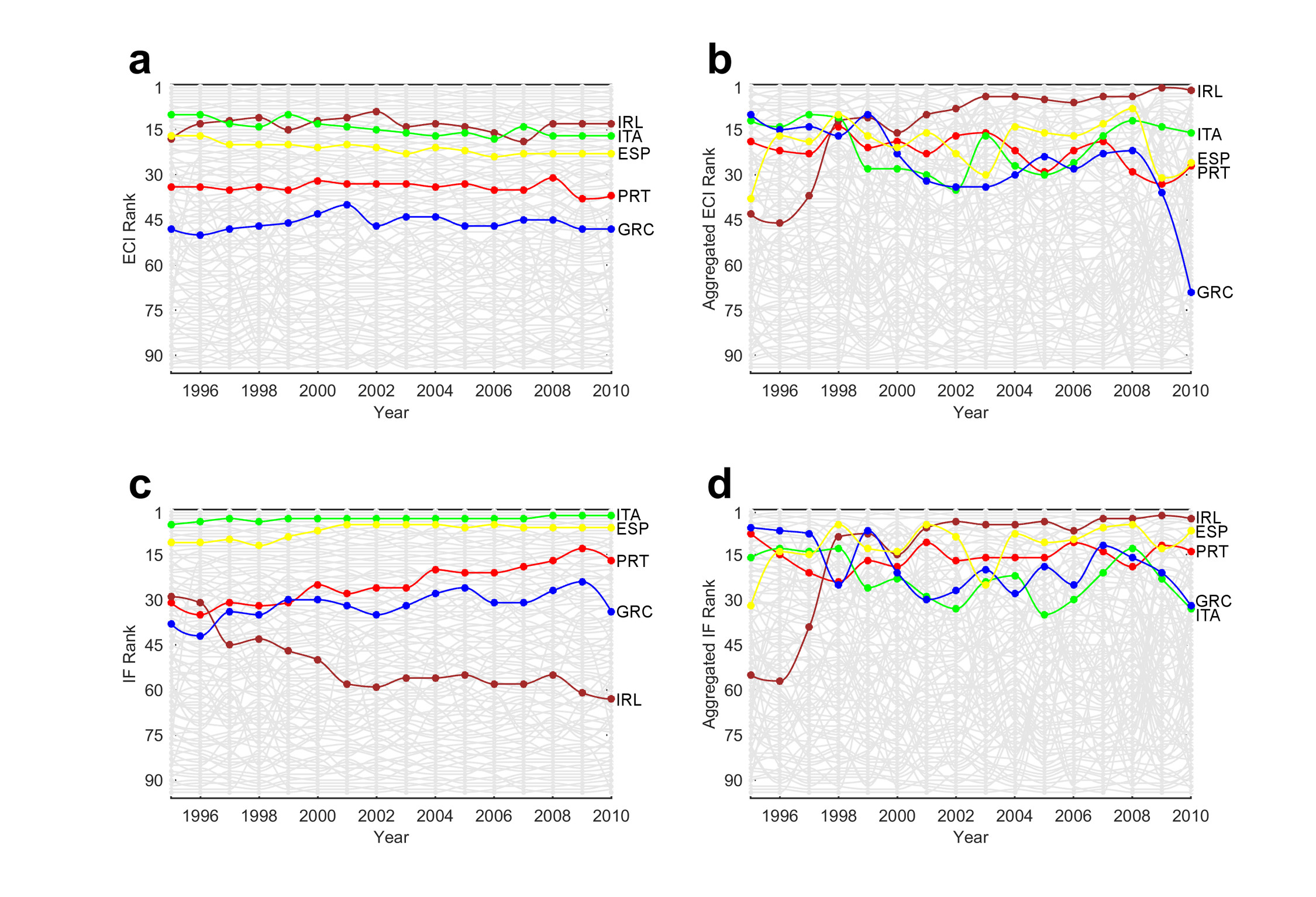}
\caption{\textbf{ Dynamics of the PIIGS nations (1995-2010).} \textbf{a)} ECI rankings movement  \textbf{b)} Same for aggregated ECI. \textbf{c)} IF rankings movement \textbf{d)} Same for aggregated IF. \textbf{a-d} Three digit iso codes are used as country abbreviations.}   \label{fig:piigs}
\end{adjustwidth}
\end{figure}

Fig. \ref{fig:piigs} illustrates the disaggregated and aggregated dynamics of the PIIGS countries. Even though the aggregated ECI and IF dynamics (Fig. \ref{fig:piigs}b and Fig. \ref{fig:piigs}d) moderately match their disaggregated counterparts (Fig. \ref{fig:piigs}a and Fig. \ref{fig:piigs}c) with the "partially" stable (ranking) movements of Portugal, Italy and Spain, there are some significant differences in the movements of the other countries. For example, Greece has very big fluctuations which lead to a dramatic drop in its complexity in the period of 2009-10 (if we examine only the aggregated IF, the same conclusions hold for Italy). Additionally, in contrast to \cite{Cristelli-2013}, our aggregated rankings reveal that Ireland has greatly transformed its productive structure and is presently one of the most complex economies. We believe that these dynamics explain better the long term economic expectations of the PIIGS nations than the projections based solely on goods data. As a matter of fact, Ireland was the first country to exit its bailout program by the end of 2013 \cite{Ireland}, and in the following years experienced significant economic growth. On the other hand, the economy of Greece did not experience the same improvements and, as of 2016, is yet to make its full recovery; the country was even reclassified as an emerging market by some equity index providers \cite{Greece}.

It should be emphasized that we do not dispute the fact that the PIIGS crisis may be attributed, at least in part, to the structural problems in the countries' economic systems and the inflexibility of their policies. In addition, as the authors in \cite{Cristelli-2015} argue, different regimes may exist for the economic complexity leading to the there introduced Selective Predictability model. Instead, we provide an additional explanation for the recent economic situation in these countries. In particular, we address the re-emergence of Ireland's economy and the ongoing stagnation of Greece. Unlike Ireland, which over the years managed to include more complex services to its product portfolio, Greece maintained its service productivity only in industries such as shipping and tourism. These services, respectively included under EBOPS 205: Transportation and EBOPS 236: Travel in our aggregated classification, not only rank lower on the product complexity scale, but also show to be especially sensitive to financial crises.

\section*{Conclusion}
\label{Conclusion}

We investigated the impact of services on the countries' productive structure in the context of the recently developed theory of economic complexity. By including service data and aggregating goods data (to make it comparable to the service data), we introduced aggregated indices as measures for the complexity of products (here denoting both goods and services) and their producers (i.e. countries). 

Similar to previous works dealing with economic complexity, we described the relations between countries and products they export via a bipartite network (graph), along with the thereby associated country-product (adjacency) matrix, with the difference that the adjacency matrix was now derived from aggregated goods and service data. The complexity measures were obtained by using two standard algorithms for evaluating complexity - the (linear) method of reflections (MR) due to Hidalgo and Hausmann, and the (nonlinear) fitness-complexity method (FCM) due to Tachella et al. While the application of the MR method to the aggregated goods and service data was straightforward, the direct application of the FCM method faced the problem of convergence of some of the (country) fitness and (product) complexity scores to zero values, effectively distorting the ranking picture. This came as result of the "unfavorable shape" of the country-product matrix (as derived from real trade data), which is along the lines of the results in \cite{Pugliese-2014} and \cite{Wu-2016}, which relate the convergence of the fitness and complexity scores obtained from the FCM method to the shape of the underlying country-product matrix. To deal with this issue, we proposed a modification of the fitness-complexity method, which may be interpreted as a second order approximation of the original FCM. We thereby proved that, with the modification in place, the complexity indices (both for countries and products) converge to values strictly greater than zero.    

Our analysis showed that, although some information was inevitably lost in the process of aggregation, the inclusion of services in the model provided new insights for both countries and products. Importantly, the statistically significant relationship between the complexity measures and the economic growth remained even with the aggregated product nomenclature (including goods and services). Nonetheless, these aggregated metrics are not as good explanatory variables of growth as their disaggregated versions. Indeed, although we add service information, due to the aggregation we now extract less information about the productive capabilities embedded in the goods. Regardless of that, the aggregation still allows us to correctly grasp the differences between the complexity of goods and services. Importantly, we discovered that services are, on average, more complex than goods, with some services topping the overall product rankings. Given the premise of a capability-driven productive structure, we conclude that sophisticated services require far more capabilities as a means to be produced, as compared to goods. 

Next, we investigated the impact that the inclusion of services has on the overall structure of the country rankings. We concluded that the addition of the service exports in the model increased the complexity of economies with developed service sector. Following standard interpretation which relates economic complexity with the potential for future growth, conjectured that the economies whose export basket, includes (sophisticated) services have bigger long-term growth potential as opposed to economies centered on export of goods.

However, we also discovered that the newly obtained aggregated country dynamics experience bigger changes over the years when compared to the disaggregated ones. While this effect may be attributed in part to the aggregation of goods data, the large volatility of the rankings after 2008 suggests that the Financial Crisis of 2007-08 and the following recession had a drastic impact on the country and product complexity scores. As a result, it may be argued that the aggregated metrics are more susceptible to (financial) economic crises (as compared to their disaggregated version which is based on disaggregated goods data), resulting in more frequent but also bigger changes in the rankings. One interpretation of this phenomenon would be that in light of financial crises, most goods and sophisticated services (such as financial, which are usually the most complex services according to our model)  are subject to a major decrease in trade, thus significantly affecting complexity indices and country rankings.  In addition, it may be argued that the restrictive trade measures introduced in the countries during the recession affect services differently (as compared to goods). Nevertheless, the investigated country dynamics suggests that the countries with sophisticated service sector succeeded to exit recession relatively quickly, which also speaks for their (relative) robustness on long term. This is true for example for the UK and the USA economy, and in particular, for Ireland which was the first country to exit its bailout program by the end of 2013, and experienced a significant economic growth afterwards.

Lastly, by producing a country specific analysis, we argued that the model with services yields economic complexity measures which, for some countries, are more consistent with their "perceived" economic situation, as otherwise suggested by an analysis performed solely on disaggregated exported goods. Examples include the countries with developed service sectors (USA, UK and Estonia) and, in particular, the PIIGS countries where these metrics provide additional information in light of their Sovereign-Debt crisis.

A more detailed analysis is certainly needed to further uncover the effects of services on the complexity structure of the countries. This, however, requires a highly accurate and disaggregated service data which is currently not available. Another fruitful research direction would be to fully exploit the concept which relates products (goods and services) with the capabilities required for their production. Building an explanatory model for the underlying capabilities would bring novel insights about the relations between products in the product space and, importantly, about the productive knowledge embedded in a country's economy. This is a subject of our current work.

\section*{Supporting Information}

\paragraph*{Supporting Information S1.}
\label{S1}

Auxiliary results serving to assess the robustness of the results presented in the manuscript:
\begin{enumerate}
\item  Economic Complexity as Indicator of Future Growth: Robustness Check
\item  Alternative RCA representation/Concatenated metrics
\end{enumerate}


\end{document}


\begin{flushleft}
{\Large
\textbf\newline{Supporting Information S1: The Impact of Services on Economic Complexity: Service Sophistication as Route for Economic Growth}
}
\newline
\\
Viktor Stojkoski\textsuperscript{1},
Zoran Utkovski\textsuperscript{1,3},
Ljup\v{c}o Kocarev\textsuperscript{1,2,*}
\\
\bigskip
\bf{1} Macedonian Academy of Sciences and Arts, Skopje, Macedonia
\\
\bf{2} Faculty of Computer Science and Engineering, Ss. Cyril and Methodius University Skopje, Macedonia
\\
\bf{3} Faculty of Computer Science, University Goce Del\v{c}ev \v{S}tip, Macedonia 
\bigskip


* E-mail: lkocarev@manu.edu.mk

\end{flushleft}

\section{Growth regressions: Robustness Check}

We check the robustness of our econometric results by constructing a regression in which we exclude the period fixed effect, and by making three new regressions in which we introduce the initial export of goods and services (GS) as a percent of total GDP; the initial population; and the initial value added of services as a percent of total GDP (SVD) in the model. By excluding the period fixed effect we do not allow for changes in the average value of growth over time \cite{Brooks-2014}. On the other hand, by including the aforementioned variables we account for possible effects of the trade openness, the size of the country and the magnitude of the service sector over growth. As stated in  \cite{Cristelli-2015}, for population, these variables may be interpreted as capabilities, and thus be correlated with the aggregated complexity measures. This implies that we might run into the problem of multicolinearity - failure to indicate significance of the correlated variables in the proposed model. Nonetheless, as presented in Table S1 \ref{Table:RobustReg}, the aggregated complexity measures pass the test as significant predictors of long term economic even when discounting the period effect and adding the three other variables.

 
 \begin{sidewaystable} [p!]
\begin{adjustwidth}{-0.3in}{2in}
\caption{\textbf{Growth Regressions: Robustness check.}} 
\centering
\setlength{\tabcolsep}{0.4cm}
\begin{tabular}{|l|d{3}|d{3}|d{3}|d{3}|c|d{3}|d{3}|d{3}|d{3}|} 
\hline
\multicolumn{5}{|l|}{\textbf{Economic Complexity Index}} & \multicolumn{5}{||l|}{\textbf{Intensive Fitness}} \\
\hline
\multicolumn{5}{|c|}{\textbf{Dependent Variable: Growth in GDP pc}} & \multicolumn{5}{||c|}{\textbf{Dependent Variable: Growth in GDP pc}} \\
\multicolumn{5}{|c|}{\textbf{1988-1998, 1998-2008}} & \multicolumn{5}{||c|}{\textbf{1988-1998, 1998-2008}} \\
\hline
\multicolumn{1}{|c|}{\textbf{Variable}} &\multicolumn{1}{|c|}{(\rom{1})} &\multicolumn{1}{|c|}{(\rom{2})} &\multicolumn{1}{|c|}{(\rom{3})} &\multicolumn{1}{|c|}{(\rom{4})} & \multicolumn{1}{||c|}{\textbf{Variable}} &\multicolumn{1}{|c|}{(\rom{1})} &\multicolumn{1}{|c|}{(\rom{2})} &\multicolumn{1}{|c|}{(\rom{3})} &\multicolumn{1}{|c|}{(\rom{4})} \\
\hline
\multicolumn{1}{|l|}{Income per capita, logs} & -0.044^{**} & -0.050^{***}& -0.037^{**} & -0.026 &\multicolumn{1}{||l|}{Income per capita, logs} & -0.053^{***} & -0.053^{***} & -0.041^{***} & -0.032 \\\hline
& (0.014) & (0.016) & (0.015) & (0.023) &\multicolumn{1}{||l|}{} & (0.014) & (0.016) & (0.015) & (0.020) \\\hline
\multicolumn{1}{|l|}{Increase in NR exports} & 0.453^{***} & 0.322^{*}& 0.329^{**}& 0.302^{*}& \multicolumn{1}{||l|}{Increase in NR exports} & 0.422^{***} & 0.306^{*} & 0.316^{*} & 0.293^{***} \\\hline
& (0.160) & (0.174) & (0.166) & (0.176) & \multicolumn{1}{||l|}{} & (0.078) & (0.171) & (0.162) & (0.086) \\\hline
\multicolumn{1}{|l|}{aggregated ECI }  & 0.046^{***} & 0.045^{***} & 0.036^{**}& 0.035^{**}& \multicolumn{1}{||l|}{aggregated IF, logs } & 0.060^{***} & 0.048^{***} & 0.040^{**} & 0.049^{**}\\\hline
& (0.014) & (0.015) &(0.015) & (0.018) & \multicolumn{1}{||l|}{} & (0.017) & (0.017) & (0.019) & (0.021) \\\hline
\multicolumn{1}{|l|}{COI} & 0.005^{***}& 0.004^{**}& 0.005^{**} & 0.004^{**}& \multicolumn{1}{||l|}{COI}& 0.005^{**} & 0.004^{**} & 0.004^{**} & 0.004^{**} \\\hline
& (0.002) & (0.002) & (0.002) &(0.002) & \multicolumn{1}{||l|}{} & (0.002) & (0.002)& (0.002)& (0.002) \\\hline
\multicolumn{1}{|l|}{Export of GS} && 0.001^{**} &&& \multicolumn{1}{||l|}{Export of GS}&& 0.001^{**} &&\\\hline
&& (0.000) &&& \multicolumn{1}{||l|}{} && (0.000) &&\\\hline
\multicolumn{1}{|l|}{Population, logs} &&& 0.010 &&\multicolumn{1}{||l|}{Population, logs}& & & 0.009 &  \\\hline
&&& (0.014) && \multicolumn{1}{||l|}{}&  & &(0.015) & \\\hline
\multicolumn{1}{|l|}{SVD, logs}&&&& -0.092 & \multicolumn{1}{||l|}{SVD, logs}&&&& -0.102 \\\hline
&&&& (0.129) & \multicolumn{1}{||l|}{}&&&& (0.093) \\\hline
\multicolumn{1}{|l|}{Constant} & 0.650^{***}&0.675^{***}&0.420 &0.447^{*}& \multicolumn{1}{||l|}{Constant} & 0.758^{***} & 0.728^{***} & 0.508 & 0.519^{**}\\\hline
& (0.122) & (0.147) & (0.289) & (0.023) &\multicolumn{1}{||l|}{} & (0.131) & (0.148) & (0.322) & (0.219) \\\hline
\multicolumn{1}{|l|}{Observations} &\multicolumn{1}{|c|}{210} &\multicolumn{1}{|c|}{201} &\multicolumn{1}{|c|}{210} &\multicolumn{1}{|c|}{178} & \multicolumn{1}{||l|}{Observations} &\multicolumn{1}{|c|}{210} &\multicolumn{1}{|c|}{201} &\multicolumn{1}{|c|}{210} &\multicolumn{1}{|c|}{178}\\\hline
\multicolumn{1}{|l|}{$R^{2}$} & 0.178 & 0.240 & 0.243& 0.242& \multicolumn{1}{||l|}{$R^{2}$} & 0.196 & 0.242 & 0.244 & 0.252 \\\hline
\multicolumn{1}{|l|}{Year FE} &\multicolumn{1}{|c|}{No} &\multicolumn{1}{|c|}{Yes} &\multicolumn{1}{|c|}{Yes} &\multicolumn{1}{|c|}{Yes} & \multicolumn{1}{||l|}{Year FE} &\multicolumn{1}{|c|}{No} &\multicolumn{1}{|c|}{Yes} &\multicolumn{1}{|c|}{Yes} &\multicolumn{1}{|c|}{Yes} \\

\hline
\end{tabular}
\begin{flushleft}
For the regressions without fixed effects, ordinary standard errors are shown in parentheses. For all other regressions standard errors clustered by cross-section are shown. ***$p<0.01$, **$p<0.05$, *$p<0.1$

\end{flushleft}

\label{Table:RobustReg}
\end{adjustwidth}
\end{sidewaystable}

\section{Alternative RCA representation/Concatenated metrics}

Production of raw materials, manufactured goods and services are three separate processes of the economy (equivalent to the primary, secondary and tertiary sector of the economy). One may argue, due to often having vanishing fixed costs, that services have distinct export features, when compared to manufactured goods and raw materials whose cost of production can more easily be attributed to their corresponding export values. In this sense, export might no longer be an adequate presentation of the internal structure of the productive system of a country. In addition, as services may be a dominant part of some economic systems, their inclusion in the model might provide bias towards service-oriented economies. On the other hand, in spite of the potential drawbacks, it still may be argued that the concept of Revealed Comparative Advantage \cite{Balassa-1964} is structured in a way that it still accounts for the varying export costs and export volumes among products (be it services and goods). With the aim to investigate these issues to more detail, here we consider an alternative representation where the RCA indices for services are calculated separately from the RCA indices for goods:
\begin{align*}
 \label{eq:RCAconc}
RCA_{is} = \frac{E_{is}/\sum_s E_{is}}{\sum_i E_{is}/\sum_{i,s} E_{is}} \\
\\
RCA_{jg} = \frac{E_{jg}/\sum_g E_{jg}}{\sum_j E_{jg}/\sum_{j,g} E_{jg}} \\
\end{align*}
where $s\in\{1,\ldots, S\}$ and  $g\in\{1,\ldots, G\}$, with $S$ and $G$ denoting the number of services, respectively goods, in the model based on aggregated goods data. Afterwards, these indices are combined to construct the $M$ matrix, from which "concatenated complexity measures" (i.e. ECI/linear PCI and Fitness/nonlinear PCI) are derived.  

It can be argued that with this representation one \textit{effectively} "divides" the economic system of a country into two sectors - services and goods. This may bias the results because when constructing the $M$ matrix equal weights are given to the overall exports of services and goods, thus potentially failing to acknowledge the initial differences in the nature and magnitude of their production. Moreover, it can be argued, that with this we slightly move away from the notion of Economic Complexity, since we implicitly assume that capabilities diffuse separately between goods and services (under the premise of a capability-driven interpretation of the country's productive structure).

In Table S1 \ref{Table:ConcReg} we reproduce the growth regressions with the concatenated complexity metrics. Column \rom{1} states that the concatenated diversity is not a significant predictor of growth, whereas from columns \rom{2} and \rom{4} we conclude that, on $5\%$ level, the concatenated ECI and the concatenated IF significantly explain long term growth. When the models are compared, the the concatenated IF clearly outperforms the concatenated diversity, while the insignificance of the later in its individual model leads to insignificance of the model where we compare it to the concatenated ECI. As a means to infer which model performs better (whether the model with the concatenated ECI or the concatenated diversity) we must opt for another method for choosing between variables, such as the Information criterions. We estimated the Hannan-Quinn, Akaike and the Bayesian Information Criterion (not shown here, but available in the workfile and/or by request) and all of them prefer the model with the concatenated ECI. 

\begin{table} [b!]
\begin{adjustwidth}{0in}{0in}
\caption{\textbf{Concatenated Complexity Metrics and Growth}} 
\centering
\setlength{\tabcolsep}{0.6cm}
\begin{tabular}{|c|d{3}|d{3}|d{3}|d{3}|d{3}|d{3}|} 
\hline
\multicolumn{6}{|c|}{\textbf{Dependent Variable: Growth in GDP pc}} \\
\multicolumn{6}{|c|}{\textbf{1988-1998, 1998-2008}} \\
\hline
\multicolumn{1}{|c|}{\textbf{Variable}} &\multicolumn{1}{|c|}{(\rom{1})} &\multicolumn{1}{|c|}{(\rom{2})} &\multicolumn{1}{|c|}{(\rom{3})} &\multicolumn{1}{|c|}{(\rom{4})} &\multicolumn{1}{|c|}{(\rom{5})} \\
\hline
\multicolumn{1}{|l|}{Income per capita, logs}  & -0.032^{**}& -0.034^{***}& -0.035^{***}& -0.040^{***}& -0.042^{***}\\\hline
 & (0.014)& (0.014) & (0.014) & (0.015) & (0.015) \\\hline
\multicolumn{1}{|l|}{Increase in NR exports}  & 0.303^{*}& 0.318^{*}& 0.318^{*}& 0.302^{*}& 0.300^{*}\\\hline
 & (0.165)& (0.167) & (0.167) & (0.164) & (0.167)\\\hline
\multicolumn{1}{|l|}{COI} & 0.003^{**} & 0.004^{**} & 0.004^{**}&0.004^{**}&0.004^{**} \\\hline
 &(0.002)& (0.002)& (0.002) &(0.002)&(0.002)\\\hline

\multicolumn{1}{|l|}{concatenated ECI}  && 0.033^{**} & 0.031 & & \\\hline
&& (0.016) & (0.019) &  & \\\hline

\multicolumn{1}{|l|}{concatenated IF, logs}&&&& 0.034^{**} & 0.070^{*} \\\hline
&&&& (0.016) & (0.040) \\\hline
\multicolumn{1}{|l|}{concatenated Diversity} & 0.008 && 0.001 & & -0.015 \\\hline
&(0.006) & & (0.008) & & (0.014) \\\hline
\multicolumn{1}{|l|}{Constant} & 0.508^{***} & 0.562^{***} & 0.562^{***} & 0.626^{***}& 0.729^{***}\\\hline
 & (0.118)& (0.126) & (0.127) & (0.138) & (0.172) \\\hline
\multicolumn{1}{|l|}{Observations} &\multicolumn{1}{|c|}{210} &\multicolumn{1}{|c|}{210} &\multicolumn{1}{|c|}{210} &\multicolumn{1}{|c|}{210} &\multicolumn{1}{|c|}{210}  \\\hline
\multicolumn{1}{|l|}{$R^{2}$}  & 0.222 & 0.232 & 0.232 & 0.230 & 0.234\\\hline
\multicolumn{1}{|l|}{Year FE} &\multicolumn{1}{|c|}{Yes} &\multicolumn{1}{|c|}{Yes} &\multicolumn{1}{|c|}{Yes} &\multicolumn{1}{|c|}{Yes} &\multicolumn{1}{|c|}{Yes}\\

\hline
\end{tabular}
\begin{flushleft}
 Standard errors clustered by cross-section shown in parentheses. ***$p<0.01$, **$p<0.05$, *$p<0.1$

\end{flushleft}

\label{Table:ConcReg}
\end{adjustwidth}
\end{table}

Furthermore, in Fig S1 \ref{fig:aggconc} we provide the yearly correlation between the aggregated complexity measures and the concatenated complexity measures based on aggregated data. Both, linear and nonlinear, aggregated and concatenated metrics have relatively high Spearman correlation over the years. Only after 2006 their correlation gradually declines and it ranges around 0.6 until the end of the period under investigation.

Finally, in Fig S1 \ref{fig:disconc} we show the yearly Spearman correlation between the disaggregated complexity measures and the concatenated complexity measures. These correlations are very similar to those presented in Fig 3 of the main manuscript, with only minor changes. In this figure in particular, in almost every year, and for both type of metrics (linear and nonlinear), the relations are relatively weaker. Only in the last two years (2009 and 2010), the correlation between the disaggregated and concatenated ECI does not fall as much as that for the disaggregated and aggregated ECI. On the other hand, the correlation between the disaggregated and concatenated IF is by far weaker than the one estimated for the disaggregated and aggregated IF.

The comparisons between the results presented in Fig 3 of the main manuscript and Fig S1 \ref{fig:disconc}, can serve as a quantitative indicator for the conclusion that the aggregated measures, with the apparent drawback, are still a better approximation of the productive structure embedded in the goods and in the services and, as such, better indicators for the complexity of countries and products in the economic complexity terminology.

 \vfill
  \begin{center}    
      \includegraphics[width=11cm]{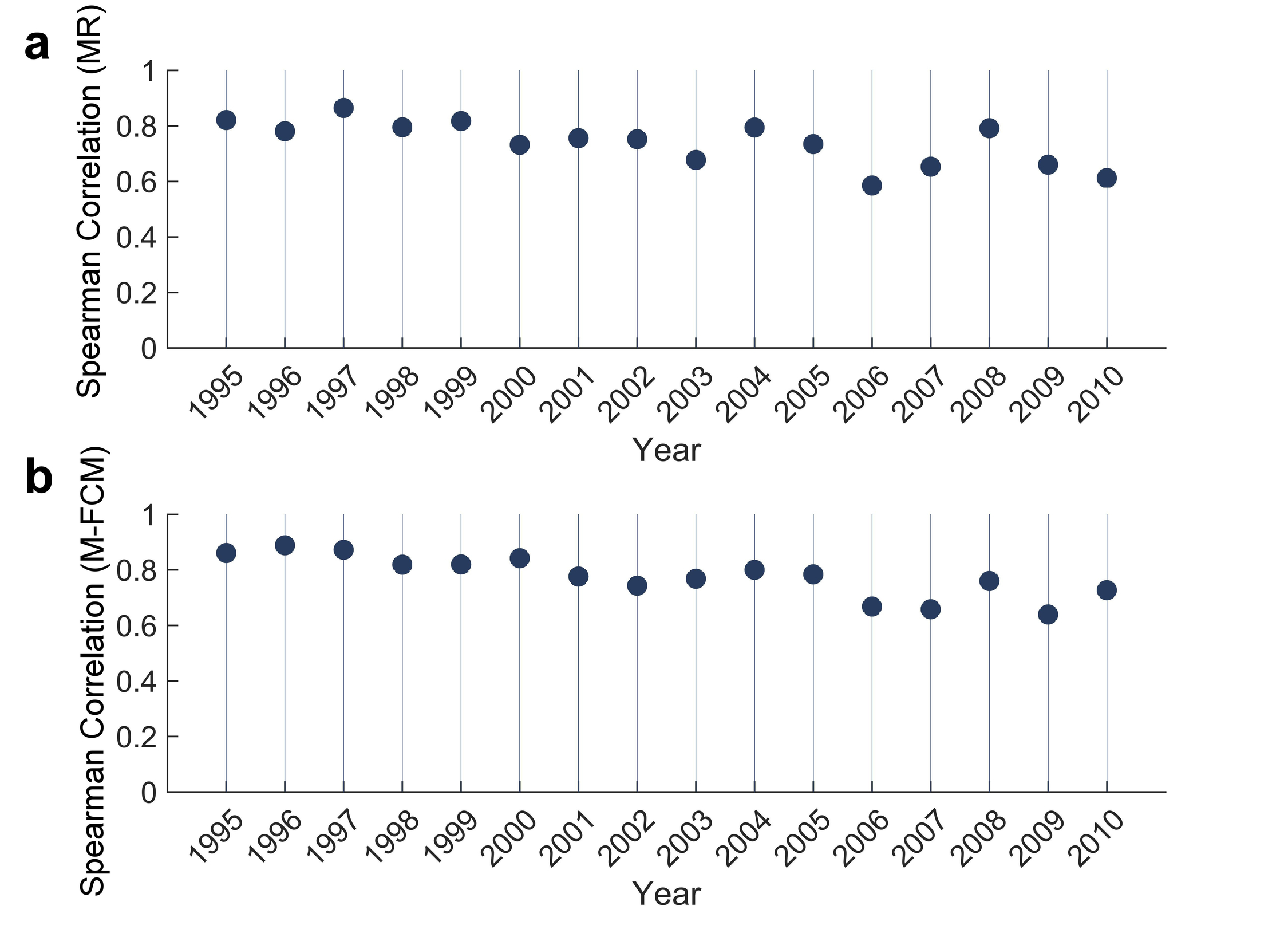}
  \end{center}
  \vfill
\begin{figure}[h!]
\caption{\textbf{Correlation between the aggregated and concatenated metrics}. \textbf{a)} Yearly Spearman correlation between the aggregated and concatenated ECI (estimated through MR). \textbf{b)} same as  \textbf{a)} for the aggregated and concatenated IF (estimated through M-FCM). \label{fig:aggconc}}
\end{figure}

 \vfill
  \begin{center}    
    \includegraphics[width=11cm]{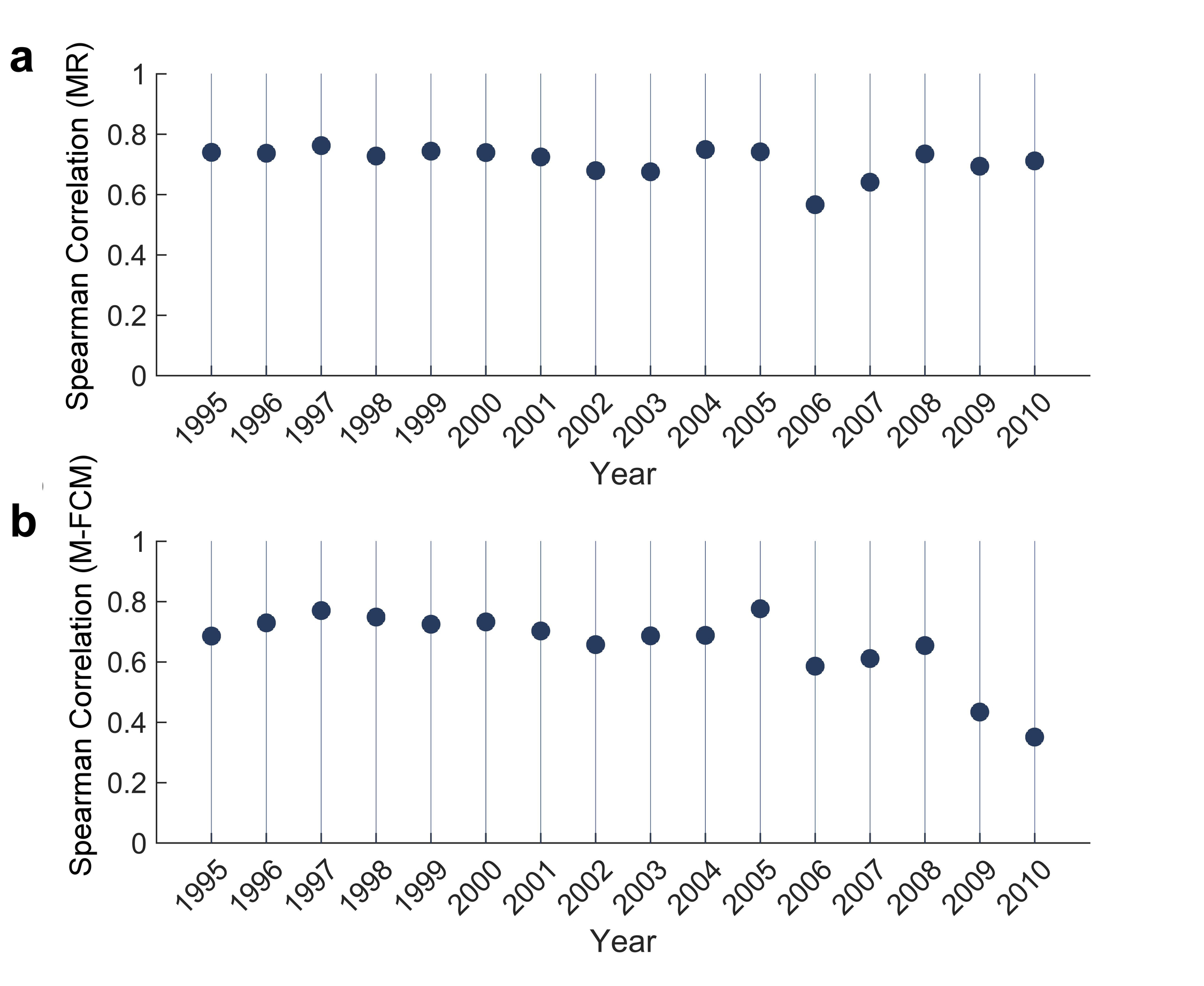}
  \end{center}
  \vfill
\begin{figure}[h!]
\caption{\textbf{Correlation between the disaggregated and concatenated metrics}. \textbf{a)} Yearly Spearman correlation between the disaggregated and concatenated ECI (estimated through MR). \textbf{b)} same as  \textbf{a)} for the disaggregated and concatenated IF (estimated through M-FCM). \label{fig:disconc}}
\end{figure}
